\documentclass[prl,twocolumn,superscriptaddress,float,aps]{revtex4-2}
\usepackage{graphicx,amsfonts,amssymb,amsmath,hyperref,hypcap,enumerate}
\usepackage{color}
\usepackage{multirow}
\newcommand{\bsl}[1]{\boldsymbol{#1}}

\newcommand{\ii}{\mathrm{i}}

\newcommand{\U}[1]{\mathrm{U}(#1)}

\newcommand{\eqnref}[1]{Eq.\,\eqref{#1}}
\newcommand{\figref}[1]{Fig.\,\ref{#1}}

\newcommand{\refcite}[1]{Ref.\,[\onlinecite{#1}]}

\newcommand{\mat}[1]{\left(\begin{matrix}#1\end{matrix}\right)}

\newcommand{\eq}[1]{\begin{equation} #1 \end{equation}}

\newcommand{\eqa}[1]{\begin{align}\begin{split} #1 \end{split}\end{align}}
\let\oldAA\AA
\renewcommand{\AA}{\text{\normalfont\oldAA}}
\newcommand{\sgn}[1]{\text{sgn}(#1)}

\usepackage[mathscr]{euscript}

\begin{document}
	\title{Phonon helicity induced by electronic Berry curvature in Dirac materials}
	\author{Lun-Hui Hu}
	\affiliation{Department of Physics, the Pennsylvania State University, University Park, PA 16802}
	\author{Jiabin Yu}
	\affiliation{Condensed Matter Theory Center, Department of Physics, University of Maryland, College Park, Maryland 20742, USA}
	\affiliation{Department of Physics, the Pennsylvania State University, University Park, PA 16802}
	\author{Ion Garate}
	\affiliation{D\'{e}partement de Physique, Institut Quantique and Regroupement Qu\'{e}b\'{e}cois sur les Mat\'{e}riaux de Pointe, Universit\'{e} de Sherbrooke, Sherbrooke, Qu\'{e}bec, Canada J1K 2R1}
	\author{Chao-Xing Liu}
	\email{cxl56@psu.edu}
	\affiliation{Department of Physics, the Pennsylvania State University, University Park, PA 16802}
	\begin{abstract}
		In two-dimensional insulators with time-reversal (TR) symmetry,  a nonzero local Berry curvature of low-energy massive Dirac fermions can give rise to nontrivial spin and charge responses, even though the integral of the Berry curvature over all occupied states is zero.
		In this work, we present a new effect induced by the electronic Berry curvature.
		By studying electron-phonon interactions in BaMnSb$_2$, a prototype two-dimensional Dirac material possessing two TR-related massive Dirac cones, we find that the nonzero local Berry curvature of electrons can induce a phonon angular momentum. The direction of this phonon angular momentum is locked to the phonon propagation direction, and thus we refer to it as ``phonon helicity", in a way that is reminiscent of electron helicity in spin-orbit-coupled electronic systems.
		We discuss possible experimental probes of such phonon helicity.
	\end{abstract}
	\maketitle

	{\it Introduction -}
	Berry phase and Berry curvature play a vital role in various branches of physics\cite{berry1984}, owing to their deep connection to gauge field theories and differential geometry\cite{Simon1983,xiao2010}.
	In condensed matter systems with broken TR symmetry, the integration of the  electronic Berry curvature over all the occupied states in the first Brillouin zone underpins a number of Hall-related phenomena, including the quantum Hall and quantum anomalous Hall effect \cite{Klitzing1980,Thouless1982,Haldane1988,Chang2013}, (intrinsic) anomalous Hall effect \cite{Nagaosa2010} and orbital magnetic moments \cite{xiao2005,xiao2006}.
	 Similarly, the Berry phase and curvature of nonelectronic quasiparticles such as phonons and magnons lead to a thermal Hall effect in insulating crystals with broken TR\cite{zhang_prl_2019,takahashi_prl_2016,park_prb_2019,lu_nl_2010,sun_prb_2020,mead_jcp_1979}.
	In a TR invariant system, the integral of Berry curvature over all the occupied states vanishes.
	Nevertheless, nonvanishing Berry curvature can still appear locally in the Brillouin zone, inducing a variety of  electronic
	phenomena, such as the spin \cite{sinova2015} and valley Hall effects \cite{xiao2007,xiao2012,mak2014}, nonlinear Hall effect \cite{Sodemann2015,Zhang2018b,xu2018,ma2019,kang2019}, piezo-electromagnetic response \cite{Martin1972,vanderbilt2000,Vaezi2013,Droth2016,rostami2018,yu2020a,yu2020b}
	and quantized circular photo-galvanic effect in Weyl semimetals
	\cite{de2017,zhang2018a,moore2019,rees2020}.
	
	In certain two-dimensional (2D) insulators with TR symmetry, the low-energy electronic excitations are 2D massive Dirac fermions.
	In these systems, the gapped Dirac cones appear in pairs at two TR-related momenta (called two valleys below) with opposite local Berry curvatures due to TR symmetry.
	Although the total Hall conductivity vanishes, as proposed in Refs.~[\onlinecite{yu2020a}] and [\onlinecite{yu2020b}], a nonzero Hall current can be induced by dynamical strain.
	Specifically, the dynamical strain plays the role of  an artificial gauge field, dubbed ``pseudo-gauge field" in literature\cite{ilan2020}.
	One may wonder whether there are more ways (other than dynamical strain) to induce nonvanishing charge and spin response \cite{sinova2015} for the local Berry curvature in TR invariant gapped systems.
	Furthermore, one may ask whether the electronic local Berry curvature can induce nontrivial behaviors in quasiparticles other than electrons.

	In this work, we study the phonon dynamics in a prototype of 2D Dirac material, BaMnSb$_2$. 
	By combining symmetry analysis and deformation potential theory, we find that the electron-phonon (e-ph) coupling in this material has the same form as the coupling between Dirac fermions and a $\U{1}$ gauge field, meaning that phonons can also act as pseudo-gauge fields and thereby induce nonvanishing Hall currents \cite{yu2020a,yu2020b}.
	The electronic local Berry curvature can induce an out-of-plane phonon angular momentum (PAM), whose sign reverses for opposite in-plane phonon momenta. We refer to it as "phonon helicity", because it resembles the helical spin texture of spin-orbit-coupled bands in the momentum space for electronic systems.
	Specifically, the nonzero electronic local Berry curvature first introduces a self-energy correction to the phonon Green's function, which is an odd function of the phonon momentum {\bf q}, and then this self-energy correction makes the lattice vibration elliptical, eventually giving rise to nonzero PAM.
	This Berry-curvature contribution to phonon dynamics can be probed through measuring the total phonon angular momentum with a temperature gradient or through optical measurements of spatial dispersion of dielectric function.

\begin{figure}[!htbp]
		\centering
		\includegraphics[width=0.8\linewidth]{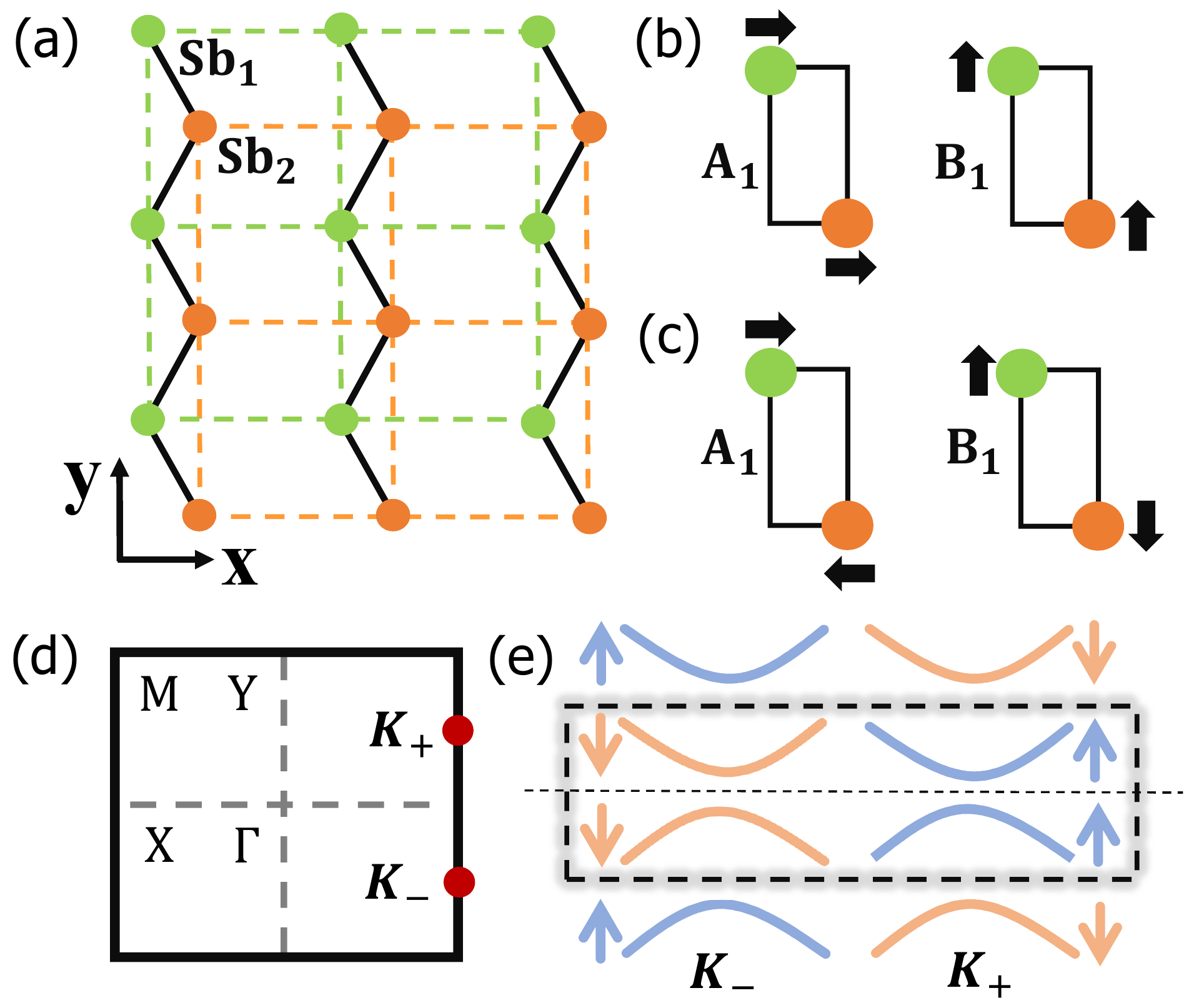}
		\caption{
			(a) Lattice structure of BaMnSb$_2$. The green and orange dots represent two Sb atoms in each unit.
			Two acoustic phonons and two optical phonons are shown in (b) and (c), respectively.
			(d) First Brillouin zone. The two red dots label the location of two valleys $K_\pm$.
			(e) Low energy electron bands at Dirac cones $K_{\pm}$. The up/down-arrow represents spin polarization direction, and our model only focuses on the bands within the box with the dashed lines.
		}
		\label{fig1}
	\end{figure}

	{\it Model Hamiltonian--}
	We take BaMnSb$_2$ as a prototype model system.
	BaMnSb$_2$ is a layered material \cite{liu2019} with alternating Ba-Sb layers and Mn-Sb layers stacked along the $(001)$ direction. Density-functional theory (DFT) calculations in Refs. [\onlinecite{farhan2014}] and [\onlinecite{liu2019}] suggest that the electronic bands near the Fermi energy mainly come from the $p$-orbitals of the Sb atoms in the Ba-Sb layers, while the Mn-Sb layers serve as insulating barriers that prevent the tunneling of electrons along the $(001)$ direction. 
	Consequently, the bands near the Fermi energy are almost non-dispersive along the $(001)$ direction, and thus BaMnSb$_2$ is a quasi-2D material. The unit cell in each Ba-Sb layer contains two Sb atoms, labeled as Sb$_1$ and Sb$_2$ in \figref{fig1}(a), both forming a square lattice. Distortion shifts Sb$_2$ atoms away from the center of the squares formed by Sb$_1$ atoms, resulting in a series of zig-zag chains.
	Due to the zig-zag distortion, the point group of BaMnSb$_2$ is $C_{2v}$, spanned by a two-fold rotation along the $x$-axis ($C_{2x}$) and $z$-directional ($\sigma_v(xy)$) mirror planes. Magnetic moments in the Mn-Sb layers have little influence on the low-energy bands near the Fermi energy in the Ba-Sb layer, and thus the TR symmetry is taken into account.
	
	We first describe the low-energy electronic properties. DFT calculations \cite{liu2019} indicate that the low-energy bands can be captured by two 2D massive Dirac cones at two momenta $K_\pm = (\pi/a,\pm k_{y_0} )$, noted in \figref{fig1}(d). The energy band sequence around $K_\pm$ is schematically shown in \figref{fig1}(e), in which four low-energy bands at each momentum come from the two atomic orbitals ($p_{x,y}$ orbitals) and two spin states of Sb$_1$ atoms.
	The low-energy Hamiltonian is $\mathcal{H}_s = \int d^2 k\Psi_{s,\mathbf{k}}^\dagger h_{s}(\mathbf{k}) \Psi_{s,\mathbf{k}}$ with
	\begin{align}\label{eq-ham0}
	h_s(\mathbf{k}) =  s v_0 (k_y \tau_3 + k_x\tau_1) + s m_0\tau_2,
	\end{align}
	where $v_0$ is the velocity of Dirac electrons and $m_0$ is the Dirac mass tuned by the zig-zag distortion. Here $s=\pm$ labels two valleys,
	and at each valley, the basis is labeled by $\Psi^\dagger_{s}(\mathbf{k}) = ( c_{s,p_x,\mathbf{k}}^\dagger, c_{s,p_y,\mathbf{k}}^\dagger)$, and $\tau_{i=0,1,2,3}$ are Pauli matrices for the orbital index.
	 The spin-valley locking allows us to only consider electrons with one specific spin direction around each valley at low energies \cite{liu2019}, as shown in Fig.~\ref{fig1}(e).
	Within this low-energy subspace, $\sigma_v(xy)$ corresponds to an identity matrix and does not change the 2D momentum.
	Thus, we only consider $C_{2x}$ and TR ($\mathcal{T}$) operations henceforth.
	The Hamiltonian ~\eqref{eq-ham0} transforms as $C_{2x}h_s(k_x,k_y)C_{2x}^{-1} = h_{-s}(k_x,-k_y)$ and $\mathcal{T}h_s(\mathbf{k}) \mathcal{T}^{-1}=h_{-s}(-\mathbf{k})$, where the symmetry representations are $C_{2x}=\tau_3$ and $\mathcal{T}=\tau_0\mathcal{K}$ ($\mathcal{K}$ stands for complex conjugation). Within one valley, we only have the combined symmetry $C_{2x}\mathcal{T}=\tau_3\mathcal{K}$.

	For the phonons, we focus on the in-plane vibrational modes, which are decoupled from the out-of-plane modes due to the opposite parities under $\sigma_v(xy)$. There are two Sb atoms in one unit cell and each atom can vibrate in the $x$ or $y$ direction. Thus, we have in total four phonon modes, two acoustic and two optical, which are classified according to the $C_{2v}$ point group, as shown in table I of Sec.~I.A in Supplemental Material (SM)~\cite{sm2021}.
	Specifically, the lattice vibrations along the $x$ and $y$ directions are decoupled at ${\bf q}=0$ due to the opposite signs under $C_{2x}$;
	the vibration patterns of these four modes are schematically shown in \figref{fig1}(b) and (c). The displacement vectors for these modes 
	are labelled by $\vec{u}_{A_1}^{a}, \vec{u}_{B_1}^{a}, \vec{u}_{A_1}^{o}, \vec{u}_{B_1}^{o}$, where $a/o$ are for acoustic and optical phonon modes and $A_1, B_1$ are the corresponding irreducible representations ($C_{2x}\vec{u}_{A_1}^{a/o}=\vec{u}_{A_1}^{a/o}$ and $C_{2x}\vec{u}_{B_1}^{a/o}=-\vec{u}_{B_1}^{a/o}$). These four phonon modes can also be obtained from the dynamical matrix for this system, as discussed in Sec.~I.B of SM.

	The e-ph coupling takes the general form
	\begin{align}\label{eq-e-ph-Ham}
	\mathcal{H}_{e-ph}^{\eta} = \sum_{\mathbf{k},\mathbf{q}}\sum_{\lambda, s,\alpha,\beta} \mathbf{g}^{\eta,\lambda,s}_{\alpha,\beta}(\mathbf{q}) Q_{\eta,\lambda,\mathbf{q}} c_{s,\alpha,\mathbf{k}} ^\dagger c_{s,\beta,\mathbf{k}-\mathbf{q}},
	\end{align}
	where  $Q_{\eta,\lambda,\mathbf{q}}=(b_{\mathbf{q},\lambda}+b^\dagger_{-\mathbf{q},\lambda})/\sqrt{2M\omega_\lambda}$ labels the phonon displacement operator for the phonon mode $\vert \eta,\lambda\rangle$
	($\eta$ for acoustic/optical and $\lambda$ for $A_1$/$B_1$);
	 $M$ is the mass of Sb atoms;
	$b_{\mathbf{q},\lambda}$ is a bosonic operator that annihilates a phonon mode $\lambda$ with momentum $\mathbf{q}$;
	$\alpha,\beta$ represent $p$-orbitals and $\mathbf{g}^{\eta,\lambda,s}_{\alpha,\beta}(\mathbf{q})$ is the element of the e-ph coupling vertex, which satisfies the Hermitian condition $\mathbf{g}^{\eta,\lambda,s}(\mathbf{q})=[\mathbf{g}^{\eta,\lambda,s}(-\mathbf{q})]^\dagger$.
	The detailed form of $\mathbf{g}^{\eta,\lambda,s}$ can be derived from the deformation theory combined with the ${\bf k\cdot p}$ theory, as shown in Secs. II and III of SM~\cite{sm2021}.
	Here we only focus on long-wavelength optical phonons (dropping the index $\eta$ below) and keep the spin-conserving intra-valley scattering process. Thus, we obtain
	\begin{equation}
	\mathbf{g}^{A_1,s} = g_0 \tau_0 + s g_2 \tau_2 + g_3\tau_3;\,\, \mathbf{g}^{B_1,s} = g_1\tau_1 ,
	\label{eq-g-terms}
	\end{equation}
	to the lowest order in ${\bf q}$, where $g_{i}$'s are real and material-dependent parameters\cite{abrikosov2012,mahan2013}.
	The e-ph coupling vertex matrices satisfy the relations $C_{2x}\mathcal{T} \mathbf{g}^{A_1,s} \mathcal{T}^{-1}C_{2x}^{-1} = \mathbf{g}^{A_1,s}$ and $C_{2x}\mathcal{T} \mathbf{g}^{B_1,s} \mathcal{T}^{-1}C_{2x}^{-1} = -\mathbf{g}^{B_1,s}$.
	Comparing \eqnref{eq-g-terms} with \eqnref{eq-ham0} (see Sec.~II.C of SM~\cite{sm2021}),
	one can see that $g_0$, $g_1$ and $g_3$ terms lead to a pseudo-gauge field
	\begin{align}
		{A}_{pse,\mu}^s =(g_0Q_{A_1}, s\tfrac{g_1}{v_0}Q_{B_1}, s\tfrac{g_3}{v_0}Q_{A_1})_{\mu},
		\end{align}
		where $e=1$ is assumed and $\mu=0,1,2$.
		The $g_2$ term acts as a correction to Dirac mass.
	Taking into account both valleys, we note that $\tfrac{g_1}{v_0}Q_{B_1}$ and  $\tfrac{g_3}{v_0}Q_{A_1}$ couple to electrons oppositely at two valleys, while the coupling between $g_0Q_{A_1}$ and electrons is the same for the two valleys.
	Similar considerations apply to acoustic phonons as well (cf. Sec.~III of SM~\cite{sm2021}).

	\begin{figure}[t]
		\centering
		\includegraphics[width=0.9\linewidth]{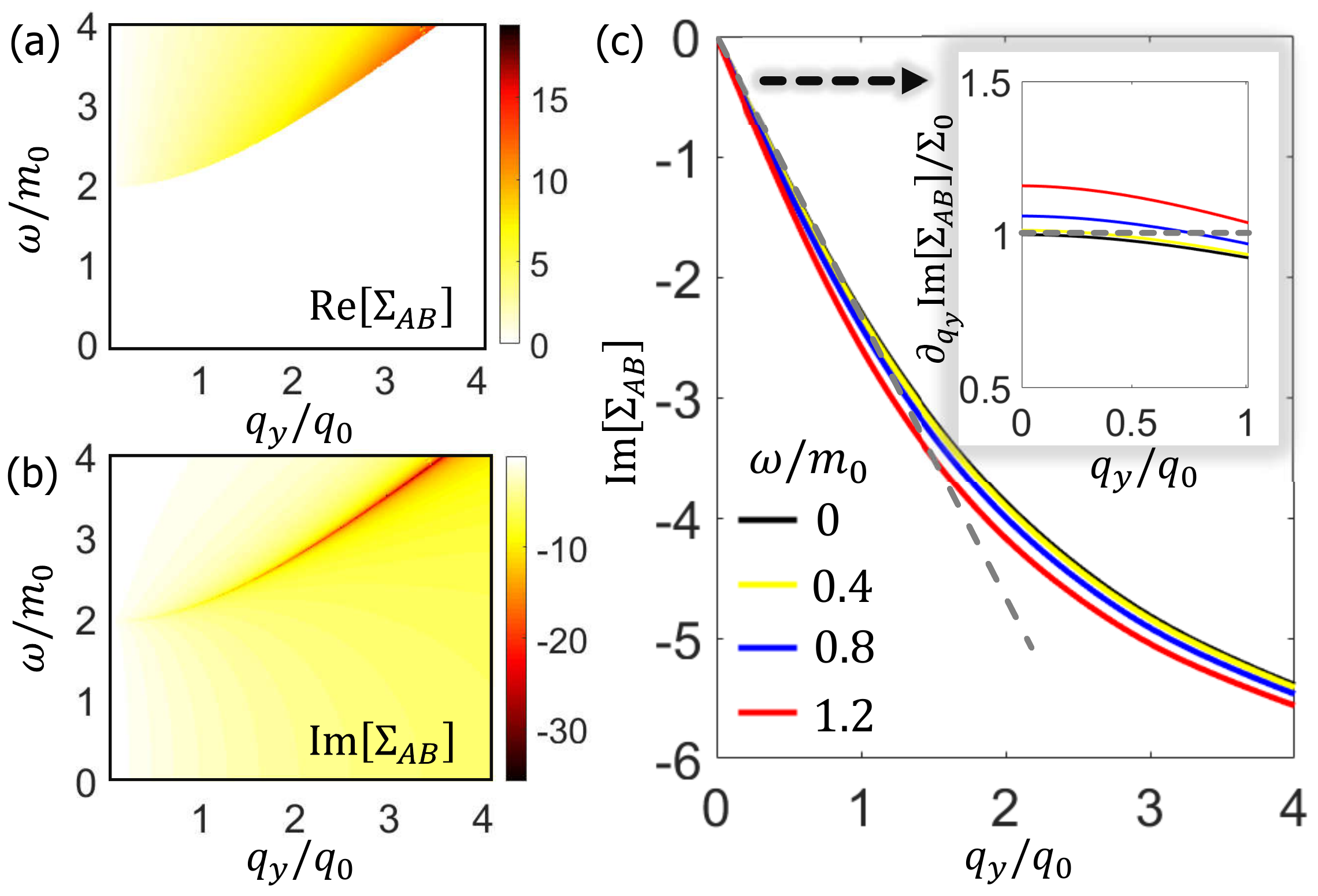}
		\caption{
			The off-diagonal phonon self-energy $\Sigma_{AB}$ induced by the electronic Berry curvature.
			We plot the real and imaginary parts of $\Sigma_{AB}$ (in meV) as a function of $\omega/m_0$ and $q_y/q_0$ in (a) and (b), respectively.
			(c) shows Im$[\Sigma_{AB}]$ as a function of $q_y$ for the frequency values $\omega/m_0=0, 0.4, 0.8, 1.2$.
			The dashed line gives the pure contribution from the Berry curvature.
			The inset shows the partial derivative of $\Sigma_{AB}$.
			The parameter values are $m_0=25$ meV, $v_0=100$ nm$\cdot$meV, $\omega_{A_1}=20$ meV, $\omega_{B_1}=30$ meV, and $g_0=80\sqrt{20}$ meV$\cdot$nm$^{-1}$, $g_1=80\sqrt{30}$ meV$\cdot$nm$^{-1}$. Also, $q_0=m_0/v_0$ and $\Sigma_0= g_0g_1\mathcal{A}N_3^+ /(v_0M\sqrt{\omega_A\omega_B})$.
		}
		\label{fig2}
	\end{figure}

	{\it Phonon self-energy from electronic Berry curvature --}
	Next, we study the optical phonon Green's function and the optical phonon spectrum renormalized by both electron-electron (e-e) and e-ph interactions.
	The bare phonon Green's function is $D_0(q,i\omega_n) = \text{Diag}\{D_{A_1},D_{B_1}\} $, where $D_{\lambda}=2\omega_{\lambda}/((i\omega_n)^2-\omega_{\lambda}^2)$ and $\omega_{\lambda}$ is the bare frequency for the $\lambda$ mode.
	 Here, $\omega_n=2\pi n/\beta$ is the Matsubara frequency and $\beta=1/k_BT$.
	Following standard formalism \cite{mahan2013}, the full phonon Green's function $D({\bf q},i\omega_n)$ can be obtained from
	$D^{-1}({\bf q},i\omega_n) = D_0^{-1}({\bf q},i\omega_n) - \Sigma({\bf q},i\omega_n)$,
	where the phonon self-energy $\Sigma=\Sigma^p+\Sigma^e$ contains both the correction from the bare e-ph interactions ($\Sigma^p$) and the additional correction due to the screened e-e Coulomb interaction ($\Sigma^e$).
	When the Fermi level $\mu$ is inside the bulk electronic gap ($\vert\mu\vert<2\vert m_0\vert$) and when $\vert\omega\vert<2\vert m_0\vert$, the $\Sigma^e$ term only renormalizes the bare dielectric constant $\epsilon_0$, the bare phonon energy $\omega_\lambda$ and the e-ph coupling strength $g_i$ (see Sec. IV of the SM~\cite{sm2021}).
	Hereafter, we only consider the contribution from the bare e-ph interaction ($\Sigma^{p}$) in \eqnref{eq-e-ph-Ham}, assuming that $\omega_\lambda$ and $g_i^o$ have been renormalized by the e-e interaction.
	Thus, the superscript $p$ is hereafter omitted.

	The diagonal components of the phonon self-energies are then found to be  $\Sigma_{AA} = (\Pi_{00} + \Pi_{33} + \Pi_{03} + \Pi_{30})/(2M\omega_A)$ and $\Sigma_{BB}=\Pi_{11}/(2M\omega_B)$, where
	\begin{align}\label{eq-self-energy-Pi-ij}
	\Pi_{ij}(\mathbf{q},i\omega_n)=\frac{g_i g_j}{\beta}\sum_{s,\mathbf{k},iq_m} \text{tr}\lbrack \tau_i G_s(\mathbf{k},iq_m) \tau_j G_s(\mathbf{k}',i\omega_n') \rbrack,
	\end{align}
	$\mathbf{k}'=\mathbf{k}+\mathbf{q}$, $i\omega_n'=i\omega_n+iq_m$, $q_m=(2m+1)\pi/\beta$, and $G_s(\mathbf{k},iq_m)=[iq_m-h_s(\mathbf{k})+\mu]^{-1}$ is the Matsubara Green's function for free electrons at valley $K_s$.
	The diagonal self-energies only provide corrections to the dispersion of the $\lambda$-phonon, and thus can be included by redefining $\omega_{\lambda}(\mathbf{q})$.

	A more interesting physical effect emerges from the off-diagonal self-energy  $\Sigma_{AB}=(\Pi_{01}+\Pi_{31})/(2M\sqrt{\omega_A\omega_B})$, where the contribution from the $g_2$ term has been neglected (see Sec.~V of SM~\cite{sm2021} for a detailed justification).
	$\Sigma_{AB}$ hybridizes A and B phonons and has a deep connection to the Berry curvature of Dirac cones.
	To see that, we first consider $\omega=0$ and treat ${\bf q}$ as a perturbation, so that $\Sigma_{AB}({\bf q},\omega=0)=\Sigma_{AB}(0)+{\bf q}\cdot (\partial_{\bf q}\Sigma_{AB})_{{\bf q}=0}+\mathcal{O}(q^2)$.
	As shown in Sec.~IV.C of SM~\cite{sm2021}, TR symmetry requires $\Sigma_{AB}(\mathbf{q},\omega)=\Sigma_{AB}^\ast(-\mathbf{q},\omega)$, while the $C_{2x}$ symmetry leads to $\Sigma_{AB}(q_x,q_y,\omega)=-\Sigma_{AB}(q_x,-q_y,\omega)$.
	Combining these two symmetry constraints together with the perturbation expansion, we find $\Sigma_{AB}({\bf q},\omega=0) \approx iq_y  (\partial_{q_y}\text{Im}[\Sigma_{AB}])_{{\bf q}=0} $.
	 We notice $\Sigma_{AB}(0)=0$ since the $A_1$ and $B_1$ modes belong to different irreducible representations, but the $q_y$-dependent term is still allowed.
	Remarkably, a direct calculation gives
	 \begin{align}\label{Eq:Berry1}
		\left( \partial_{q_y} \text{Im}[\Sigma_{AB}(q_y,0)]\right)_{q_y=0} = \frac{g_0g_1\mathcal{A}(N_3^+-N_3^-)}{2v_0M\sqrt{\omega_A\omega_B}} ,
		\end{align}
		where $\mathcal{A}$ is the sample area and the coefficient
		$N_3^s =\frac{1}{6}\epsilon^{\alpha\beta\gamma}   \text{tr}\int_{-\infty}^{\infty} d\omega' \int \frac{d^2 k}{(2\pi)^2}  G_s^{-1}  \partial_{k_\alpha} G_s G_s^{-1} \partial_{k_\beta}G_s G_s^{-1} \partial_{k_\gamma}G_s$ 
	is the Thouless-Kohmoto-Nightingale-den Nijs formula for the integral of the Berry curvature around the $s$-valley (i.e., the valley Chern number;  see Sec.~IV.D of SM~\cite{sm2021}).
	TR requires $N_3^++N_3^-=0$. Thus, to linear order in ${\bf q}$,
	 $\Sigma_{AB}({\bf q})=i g_0g_1\mathcal{A}N_3^+ q_y/(v_0M\sqrt{\omega_A\omega_B})\triangleq iq_y\Sigma_0$, and the local Berry curvature around one valley enters into the off-diagonal phonon self-energy.
	This is a key result of the present work.
	 Alternatively (see Sec.~VI of SM~\cite{sm2021}), Eq.~\eqref{Eq:Berry1} can be obtained from a Chern-Simons term that appears in the effective action for phonons after integrating out the electrons. Such derivation confirms that $\Sigma_{AB}$ results from the valley Chern numbers.
	
	The above analytical expression is derived in the limit $\omega\to 0$ and ${\bf q}\to 0$, while the $\lambda$-optical phonon has a finite frequency $\omega_\lambda$.
	To see the finite frequency behavior of $\Sigma_{AB}$, we numerically evaluate it according to \eqnref{eq-self-energy-Pi-ij}, and show the real and imaginary parts of $\Sigma_{AB}$ in \figref{fig2}(a) and (b), respectively.
	When the frequency is larger than electron energy gap ($\omega>2\vert m_0\vert$), the interband scattering of electrons from the conduction to valence bands can give rise to a large correction to the phonon self-energy.
	This is reflected by the red color lines for both $\text{Re}[\Sigma_{AB}]$ and $\text{Im}[\Sigma_{AB}]$ in \figref{fig2}(a) and (b), corresponding to the dynamical Kohn anomaly \cite{tse2008} occurring at $q= \sqrt{(\omega-m_0)^2-m_0^2}$ for $\mu=0$ (see Sec.~VII of SM~\cite{sm2021}).
	Below electron energy gap ($\omega<2\vert m_0\vert$), $\text{Re}[\Sigma_{AB}]$ is exactly zero, while $\text{Im}[\Sigma_{AB}]$ is still non-zero.
	In \figref{fig2}(c), we plot $\text{Im}[\Sigma_{AB}]$ as a function of $q_y$ for $\omega/m_0=0, 0.4, 0.8, 1.2$ separately, from which a linear-dependence on $q_y$ appears in the long-wavelength limit. The pure Berry curvature contribution extracted from Eq. (\ref{Eq:Berry1}) is indicated by the gray dashed line, which shows a good coincidence with numerical results for a small $q_y$.
	The inset in \figref{fig2}(c) shows $(\partial_{q_y} \text{Im}[\Sigma_{AB}])/\Sigma_0$ at low $q_y$, and only a small derivation from unity is found for finite $q_y$ and $\omega$.
	Therefore, our numerical simulations demonstrate that the Berry curvature contribution to the off-diagonal self-energy is dominant when the phonon frequency $\omega$ is well below electron energy gap.
	We notice that in a related material compound, SrMnSb$_2$, the optical phonon frequency has been measured to be $\sim 15\, {\rm meV}$ \cite{weber2018}, well below the expected electronic gap of $\sim 50$ meV.

	\begin{figure}[t]
		\centering
		\includegraphics[width=0.8\linewidth]{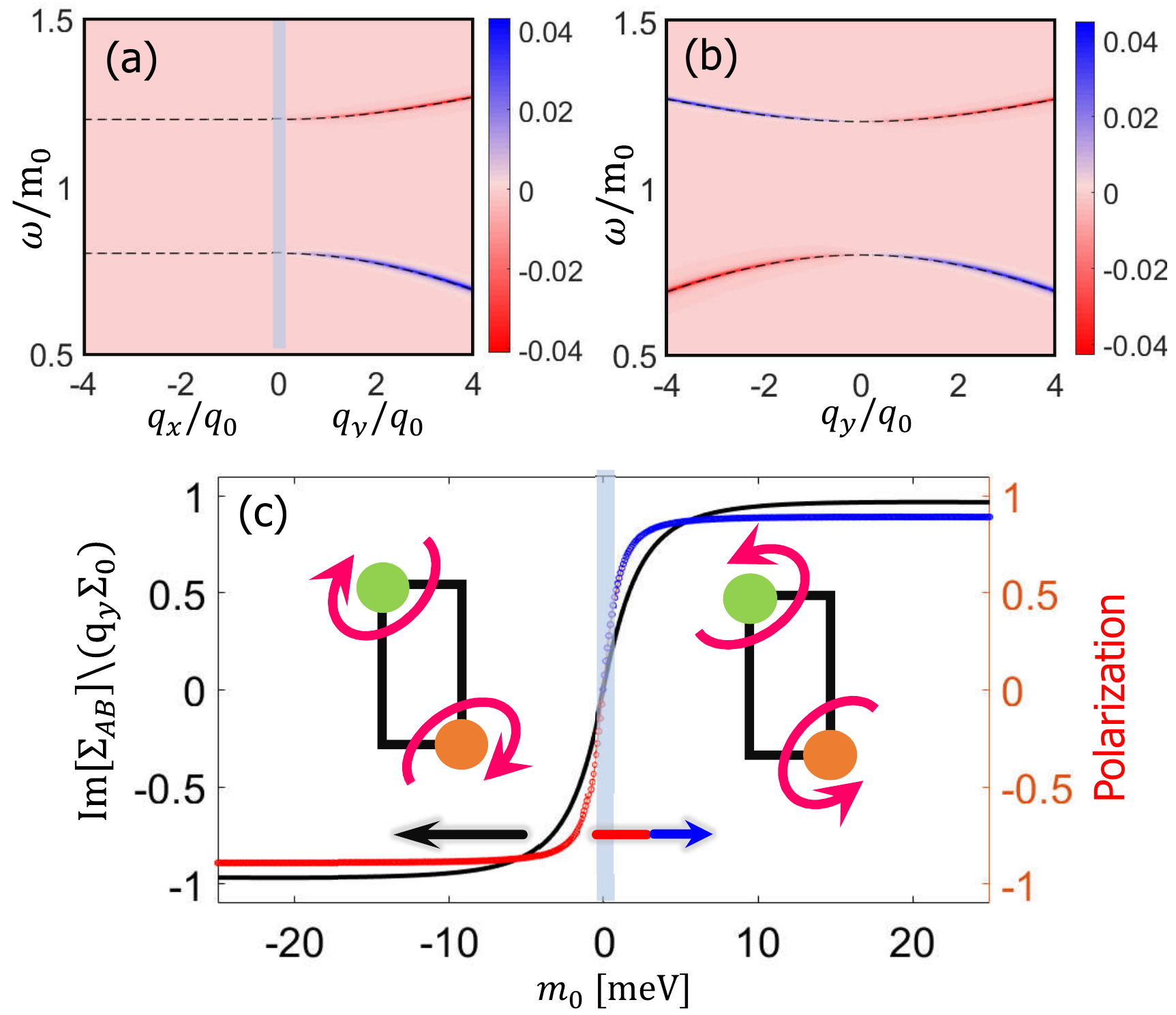}
		\caption{
			(a) and (b) show the spectrum and angular momentum of optical phonons with $m_0=25$ meV.
			The phonon dispersions are labeled by black dashed lines, while the color represents ${\cal P}$ in Eq.~\eqref{eq-phonon-AM}.
			(c) shows Im[$\Sigma_{AB}$] (black line) and the phonon circular polarization (red and blue lines) as a function of $m_0$ with $q_y=m_0/2v_0$ and $\omega=10$ meV.
			The phonon circular polarization has the same sign as ${\cal P}$.
			The inset is the schematics for the elliptical vibration of phonon modes.
			Parameters: $v_0=100$ meV$\cdot$nm, $\omega_{A_1}=20$ meV, $\omega_{B_1}=30$ meV, $g_0=80\sqrt{20}$ meV$\cdot$nm$^{-1}$, $g_1=80\sqrt{30}$ meV$\cdot$nm$^{-1}$. Also, $q_0=m_0/v_0$ and $\Sigma_0=g_0g_1\mathcal{A}N_3^+ /(v_0M\sqrt{\omega_A\omega_B})$.
		}
		\label{fig3}
	\end{figure}

	{\it Phonon helicity --}
	Next, we explore the influence of $\Sigma_{AB}$ in the phonon dynamics by studying the full phonon Green's function $D({\bf q},\omega)$. The phonon dispersion can be directly extracted from the poles of $D({\bf q},\omega)$, and is depicted by the black dashed lines in \figref{fig3}(a-b) for optical phonons.
	Since the $A_1$ and $B_1$ phonon modes describe lattice vibration along the $x$ and $y$ directions, respectively, and the off-diagonal term $\Sigma_{AB}$ that couples these two modes is purely imaginary, we expect that $\Sigma_{AB}$ can make phonon modes elliptically polarized at finite $q_y$, as schematically depicted in the inset of \figref{fig3}(c). To quantify this, we define the momentum-resolved PAM as
	\begin{align}\label{eq-phonon-AM}
	\mathcal{P}(\mathbf{q},\omega) = \text{Tr}\left[\left(\hat{P}_{+} - \hat{P}_{-} \right) \hat{A}(\mathbf{q},\omega) \right],
	\end{align}
	where $\hat{A}(\mathbf{q},\omega)=i[D(\mathbf{q},\omega)-D^{\dagger}(\mathbf{q},\omega)]/(2\pi)$  is the phonon spectral function and the projection operators $\hat{P}_{\pm} = \vert\pm\rangle \langle\pm\vert$ project the phonon modes to the left and right circular polarized basis $\vert\pm\rangle = (\vec{u}_{o,A_1}\pm i\,\vec{u}_{o,B_1})/\sqrt{2}$.
	The total PAM discussed in literature \cite{zhang2014,zhang2015,hamada2018,juraschek2019} is the thermal average of $\mathcal{P}(\mathbf{q},\omega)$ over all the phonon modes.
	We highlight two features of ${\cal P}$:
	(1) $\mathcal{P}({\bf q},\omega)=0$ when ${\bf q} || \hat{\bf x}$ but $\mathcal{P}({\bf q},\omega)\neq 0$ when ${\bf q}||\hat{\bf y}$ (see \figref{fig3}(a));
	(2) $\mathcal{P}(q_y,\omega)=-\mathcal{P}(-q_y,\omega)$, as required by TR symmetry (see \figref{fig3}(b)). The latter feature suggests the helical nature of each phonon band and gives rise to phonon helicity.
	Both features can be understood from the symmetry property of $\Sigma_{AB}$:
	(1) $\Sigma_{AB}$ vanishes along the $q_x$ direction;
	(2) $\Sigma_{AB}\propto q_y$, so that its sign reverses under $q_y\to -q_y$.
	
	More interestingly, since the local Berry curvature around one valley
	changes sign under a band inversion,
	we anticipate that the phonon helicity will also reverse its sign.
	To test this, we tune the Dirac mass $m_0$ in \eqnref{eq-ham0}, and evaluate $\Sigma_{AB}$ and $\mathcal{P}$ for the lower branch in \figref{fig3}(b) as a function of $m_0$ at the momentum $q_y=m_0/2v_0$ in \figref{fig3}(c).
	One can clearly see that both $\Sigma_{AB}$ and $\mathcal{P}$ reverse signs across the phase transition point at $m_0=0$. Thus, we conclude that an electronic topological phase transition will leave its fingerprint on PAM $\mathcal{P}$.
	
	We also notice that along the $q_y$ direction, the $B_1$ mode is longitudinal while the $A_1$ mode is transverse. Therefore, a nonzero $\Sigma_{AB}$ mixes the longitudinal and transverse modes of the in-plane phonons to form elliptical vibrational modes. The longitudinal-transverse mixing has been discussed previously in the context of the phonon Hall viscosity \cite{barkeshli2012,liu2017,heidari2019}.
	However, there is an essential difference: our system does not break TR while the phonon Hall viscosity breaks TR. Consequently, in \refcite{barkeshli2012}, the PAM is an even function of ${\bf q}$, as reflected by the $q^2$-dependent coupling term between the longitudinal and transverse modes, while in our system, the PAM is an odd function of ${\bf q}$, thereby generating phonon helicity.

	{\it Discussion and Conclusion--.}
	We have demonstrated that electronic Berry curvature in a 2D TR-invariant Dirac material can induce an off-diagonal self-energy correction to the optical phonon Green's function, giving rise to elliptical polarization of phonon modes.
	Our theory can also be directly applied to acoustic phonons.
	However, the e-ph coupling linearly depends on ${\bf q}$ for acoustic phonons
	, and thus the off-diagonal phonon self-energy  has a $q_y^3$ dependence, 
	as discussed in Sec.~VI of SM~\cite{sm2021}.
	
	Our study is qualitatively different from recent works investigating the influence of the chiral anomaly on sound velocity and attenuation in Weyl/Dirac semimetals\cite{chernodub2019,laliberte2020,sengupta2020,antebi2020,sukhachov2021}. For one thing, we focus on electrical insulators.  Despite of several early works studying the Berry phase effect on phonon dynamics \cite{lu_nl_2010,sun_prb_2020,mead_jcp_1979}, the influence of electronic band topology on the PAM and helicity have not been mentioned before.
	Furthermore, although we take BaMnSb$_2$ as an example, the proposed mechanism for generating phonon helicity might be adapted to other
	Dirac materials.
	While a careful treatment is required to verify this for each material, possible candidates include HgTe/CdTe and InAs/GaSb quantum wells\cite{bernevig2006,konig2007,liu2008,knez2011}, (LaO)$_2$(SbSe$_2$)$_2$\cite{dong2015}, boron nitride\cite{xue2011,yankowitz2012}, 1T'-WTe$_2$ \cite{wu2018,fei2017,tang2017}, and others\cite{wehling2014,wang2015,ren2016,Armitage2018}.
	
	The Berry curvature contribution to phonon helicity can be detected in the following ways.
	Due to TR symmetry, the equilibrium phonon occupation number is symmetric under $q_y\to -q_y$; the total PAM is accordingly zero. However, away from equilibrium, a temperature gradient applied along the $y$ axis breaks the $q_y\to -q_y$ symmetry and, combined with phonon helicity, leads to a nonzero total PAM. This mechanism is analogous to the Edelstein effect in electronic systems and has been discussed in Ref.~[\onlinecite{hamada2018}].
	For the optical phonon that can directly couple to light, the elliptical polarization will be reflected in the dielectric function and thus might be probed through optical measurements, as discussed in Sec.~VIII of SM~\cite{sm2021}.

	{\it Acknowledgments--.}
	We thank Zhen Bi, Shengxi Huang, Jainendra Jain, Ziqiang Mao, Yinming Shao, and Shuang Zhang for helpful discussions.
	This work at Penn State is primarily supported by a DOE grant (DE-SC0019064). L.-H. H. also acknowledges the support of the Office of Naval Research (Grant 361 No. N00014-18-1-2793). I. Garate acknowledges financial support from the Canada First Research Excellence Fund (CFREF) and the Natural Sciences and Engineering Research Council of Canada (NSERC).
	J.~Yu is supported by the Laboratory for Physical Sciences.

\bibliographystyle{apsrev4-2}
%


\clearpage
\appendix
\onecolumngrid
\setcounter{page}{1}

\begin{center}
	\bf	Supplementary material for ``Phonon helicity induced by electronic Berry curvature in Dirac materials''
\end{center}

\section{Long-wavelength phonon modes for the quasi-2D S$\text{b}$ layers}
In this section, we study the phonon modes of BaMnSb$_2$ with zigzag distortion \cite{liu2019}, with a focus on the the effective quasi-2D Sb layer (see Fig.~\ref{sm_fig1}(a)).
First, we classify the phonon modes according to the \(C_{2v}\) point group.
This group has three symmetry operators: $C_{2x}$ is the two-fold rotation along the $x$-axis, while $\sigma_v(xy)$ and $\sigma_v(xz)$ are two mirror reflections about the $xy$ and $xz$ planes, respectively:
\begin{align}
C_{2v} = \{  C_{2x},  \sigma_v(xy),\sigma_v(xz)\}.	
\end{align}
There are two Sb atoms in each unit cell (labeled by green and orange filled circles in Fig.~\ref{sm_fig1}) and each atom can move along $x$, $y$ and $z$ directions.
Thus, the total number of degrees of freedom is six, giving rise to six phonon modes.
We characterize the long-wavelength phonon modes by the displacements of the two atoms away from their equilibrium positions,  $(u_{1,x},u_{1,y},u_{1,z},u_{2,x},u_{2,y},u_{2,z})$.

\begin{table}[!htbp]
	\caption{\label{supp-tab:phonon modes} The phonon modes for the quasi-2D Sb layers, including three acoustic and three phonon modes. The corresponding lattice vibrations at $\Gamma$ point are classified by the $C_{2v}$ point group.}
	\begin{ruledtabular}
		\begin{tabular}{cc|c|c|c|c|c}
			\multicolumn{2}{c|}{Phonon modes} & $E$ & $C_{2x}$  & $\sigma_v(xy)$  &  $\sigma_v(xz)$  &  Irrep. \\ \hline
			\multirow{3}{*}{Acoustic}  &  $(1,0,0,1,0,0)$ & $+1$ & $+1$ & $+1$ & $+1$ & $A_1$ \\ \cline{2-7}
			&  $(0,1,0,0,1,0)$ & $+1$ & $-1$ & $+1$ & $-1$ & $B_1$ \\ \cline{2-7}
			&  $(0,0,1,0,0,1)$ & $+1$ & $-1$ & $-1$ & $+1$ & $B_2$ \\ \hline
			\multirow{3}{*}{Optical}   &  $(1,0,0,-1,0,0)$ & $+1$ & $+1$ & $+1$ & $+1$ & $A_1$ \\ \cline{2-7}
			&  $(0,1,0,0,-1,0)$ & $+1$ & $-1$ & $+1$ & $-1$ & $B_1$ \\ \cline{2-7}
			&  $(0,0,1,0,0,-1)$ & $+1$ & $-1$ & $-1$ & $+1$ & $B_2$ \\
		\end{tabular}
	\end{ruledtabular}
\end{table}

The results are summarized in Table~\ref{supp-tab:phonon modes}, from which we learn that the lattice vibration along the $z$-axis is odd under mirror reflection $\sigma_v(xy)$, while that along the $x$ or $y$ directions is even. Therefore, the $z$-direction mode can not couple to the electronic states around Fermi energy, which are even under $\sigma_v(xy)$.
As a result, we will only focus on the in-plane phonon modes.

\subsection{Symmetry-based classification of in-plane phonons}
\begin{figure}[!htbp]
	\centering
	\includegraphics[width=0.65\linewidth]{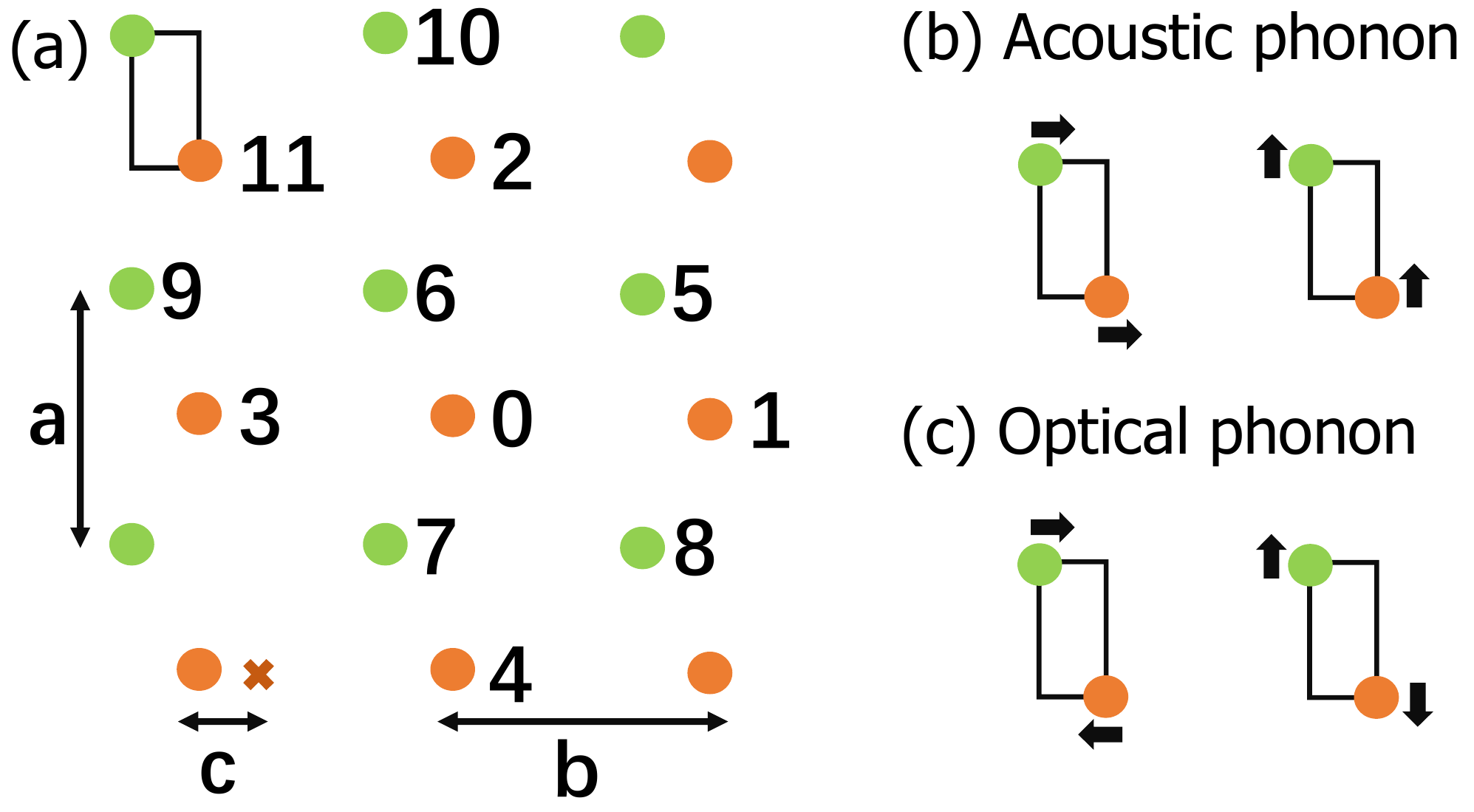}
	\caption{In (a), it shows the lattice structure for Sb layer. In (b), the acoustic phonon mode. In (c), the optical phonon mode. The black arrow indicates the relative vibration direction.
	}
	\label{sm_fig1}
\end{figure}

Once again, we only consider the effective 2D lattice structure for Sb atoms with the zigzag distorted phase of BaMnSb$_2$ (see Fig.~\ref{sm_fig1}(a)).
To study the lattice vibration in the $x-y$ plane, we ignore the mirror $\sigma_v(xy)$, which acts as an identity for the in-plane phonon modes.
In addition, $C_{2x}$ and $\sigma_v(xz)$ act in the same way on the in-plane phonon modes; hereafter only $C_{2x}$ will be taken into account for the symmetry analysis for in-plane phonon modes.
Then, effectively the point group symmetry is reduced to $C_2$.
There are in total four degree of freedom for a unit cell, i.e., one Sb with $(u_{1,x},u_{1,y})$ and the other Sb with $(u_{2,x},u_{2,y})$. Under the $C_{2x}$ rotation,
\begin{align}
C_{2x}: \begin{cases}
(u_{1,x},u_{1,y}) \to (u_{1,x},-u_{1,y}), \\
(u_{2,x},u_{2,y}) \to (u_{2,x},-u_{2,y}).
\end{cases}
\end{align}
Therefore, the reducible characters for the in-plane phonon modes are $\chi(E)=4$ and $\chi(C_{2x})=0$.
Based on the $C_2$ point group, the are two $A_1$ and two $B_1$ irreducible modes \cite{Dresselhaus_book},
which are shown in Table.~\ref{supp-tab:phonon modes}.
Here $A_1$ and $B_1$ are two 1D irreducible representations (irrep.)  in Mulliken symbols, which is taken to label the phonon modes.
Thus, there are two $A_1$-type phonon modes as well as two $B_1$-type phonon modes.
Next, we will solve the dynamical ($D$-) matrix to show the phonon modes explicitly.

\subsection{Characterization of in-plane phonons from the dynamical matrix}
In this subsection, we determine the frequencies and displacement vectors corresponding to in-plane phonon modes.
The representative Sb atoms are labeled by integer numbers shown in Fig.~\ref{sm_fig1}(a).
We treat the interatomic forces in the harmonic approximation, and we only consider the spring constants between nearest and next-nearest atoms:
\begin{align}\label{sm-eq-spring-constants}
\begin{split}
&\text{Spring constant between } \text{Sb}_0 \; \text{and}\;  \text{Sb}_{1,2,3,4} : \gamma_0,\\
&\text{Spring constant between } \text{Sb}_0 \; \text{and}\;  \text{Sb}_{6,7} : \gamma_1, \\
&\text{Spring constant between } \text{Sb}_0 \; \text{and}\;  \text{Sb}_{5,8} : \gamma_2.
\end{split}
\end{align}
Here, we have neglected the forces between neighboring Sb planes, which generally lead to vibrations along the $z$-axis.
Next, we calculate the force constant directly.
For simplicity, we take lattice constant $a=b$, and assume $c=\delta a\times\frac{a}{2}$ with $-1<\delta a<1$; $\delta a=0$ corresponds to the absence of zigzag distortion.
Since we have two Sb atoms in one unit cell, we consider them separately.

\subsubsection{Force constants for the first Sb atom}
First, let us choose the position of the Sb$_0$ atom as the origin of coordinates. The normalized coordinates for the nearest and the next-nearest atoms are
\begin{align}\label{sm-eq-neighbor-coordinates}
\begin{split}
&\text{Sb}_1: (1,0),\; \text{Sb}_2: (0,1),\; \text{Sb}_3: (-1,0),\; \text{Sb}_4: (0,-1), \\
&\text{Sb}_5: (\delta a+1,1)/A,\;  \text{Sb}_6: (\delta a-1,1)/B,\; \text{Sb}_7: (\delta a-1,-1)/B,\; \text{Sb}_8: (\delta a+1,-1)/A,
\end{split}	
\end{align}
with normalization factors $A=\sqrt{1+(\delta a+1)^2}$ and $B=\sqrt{1+(\delta a-1)^2}$.
Let us briefly review how to calculate the force constants, using Sb$_0$ as an example. The potential energy associated to the displacement of this atom is
\begin{align}
E \approx E_0 + \sum_i\sum_{\alpha,\beta=\{x,y\}} \phi_{\alpha,\beta}(0,i) {u}_{0,\alpha}  {u}_{i,\beta},
\end{align}
where $E_0$ is a constant and the index $i$ represents the nearby atoms: $i=5,6,7,8$ for the nearest neighbors and $i=1,2,3,4$ for the next-nearest neighbors (see Fig.~\ref{sm_fig1}(a)). The force constant is defined as the second derivative of the potential energy evaluated at the equilibrium position,
\begin{align}
\phi_{\alpha,\beta}(0,i) = \frac{\partial^2 E}{\partial u_{0,\alpha} \partial u_{i,\beta} } = - \frac{\partial F_\alpha} {\partial u_{i,\beta}},
\end{align}
where $ F_\alpha = \sum_i f_\alpha(0,i) $ is the total force acting on the Sb$_0$ atom along the $\alpha$-direction.
With the spring constants $\gamma(0,i)$ between Sb$_0$ and its neighbors Sb$_i$ given in Eq.~\eqref{sm-eq-spring-constants}, we have
\begin{align}
F_\alpha = \sum_{i} \gamma(0,i) e_\alpha(i) \left( \sum_{\beta} e_{\beta}(i) u_{i,\beta}  \right),
\end{align}
where $\vec{e}(i) = (e_x(i),e_y(i))$ is the unit vector along ${\bf R}_i$ (the straight line connecting Sb$_0$ and Sb$_i$), obtained from Eq.~\eqref{sm-eq-neighbor-coordinates}.
Therefore, we have the force constant
\begin{align}
\phi_{\alpha\beta}(0,i) = - \gamma(0,i) e_\alpha(i)  e_{\beta}(i).
\end{align}
In this way, we calculate all the force constants between Sb$_0$ and its neighbors (Sb$_{5,6,7,8}$ and Sb$_{1,2,3,4}$). For the nearest neighbors, we get
\begin{equation*}
\begin{aligned}[c]
\phi_{xx}(0,5) &= \phi_{xx}(0,8)=  -\gamma_2\left( \frac{\delta a+1}{A} \right)^2, \nonumber\\
\phi_{yy}(0,5) &= \phi_{yy}(0,8) = -\gamma_2\left( \frac{1}{A} \right)^2,   \nonumber \\
\phi_{xy}(0,5) &= -\phi_{xy}(0,8) = -\gamma_2\frac{\delta a+1}{A^2},
\end{aligned}
\text{      }
\begin{aligned}[c]
\phi_{xx}(0,6) &= \phi_{xx}(0,7) = -\gamma_1\left( \frac{\delta a-1}{B} \right)^2,\nonumber\\
\phi_{yy}(0,6) &= \phi_{yy}(0,7) =  -\gamma_1\left( \frac{1}{B} \right)^2,\nonumber \\
\phi_{xy}(0,6) &= -\phi_{xy}(0,7) = -\gamma_1\frac{\delta a-1}{B^2}.
\end{aligned}
\end{equation*}
The force constants for the next-nearest neighbors are
\begin{align}
\begin{split}
&\phi_{xx}(0,1) = \phi_{xx}(0,3) = \phi_{yy}(0,2) = \phi_{yy}(0,4)= -\gamma_0,
\end{split}
\end{align}
all others being zero.
Moreover, the potential energy of the crystal and the force on a given atom should be invariant under a rigid body displacement of the whole crystal.
This implies the constraint
\begin{align}
\phi_{\alpha\beta}(0,0) = -\sum_{i=1}^{8}\phi_{\alpha\beta}(0,i).
\end{align}
Therefore, we have
\begin{align}\label{sm-eq-force-conservation}
\begin{split}
\phi_{xx}(0,0) &= - \sum_{i=1}^{8}\left[ \phi_{xx}(0,i) \right] = 2\gamma_0 + 2\gamma_2\left( \frac{\delta a+1}{A} \right)^2 + 2\gamma_1\left( \frac{\delta a-1}{B} \right)^2, \\
\phi_{yy}(0,0) &= -\sum_{i=1}^{8}\left[ \phi_{yy}(0,i) \right] = 2\gamma_0 + 2\gamma_2\left( \frac{1}{A} \right)^2 + 2\gamma_1\left( \frac{1}{B} \right)^2, \\
\phi_{xy}(0,0) &= \phi_{yx}(0,0) = 0.
\end{split}
\end{align}

\subsubsection{Force constants for the second Sb atom}
Next, let us consider the second Sb atom and choose the position of the Sb$_6$ as the origin of coordinates (see Fig.~\ref{sm_fig1}(a)).
Repeating the above calculations for the force constants, we get
\begin{equation*}
\begin{aligned}[c]
\phi_{xx}(6,11) &= \phi_{xx}(6,3)=  -\gamma_2\left( \frac{\delta a+1}{A} \right)^2,\\
\phi_{yy}(6,11) &= \phi_{yy}(6,3) = -\gamma_2\left( \frac{1}{A} \right)^2, \\
\phi_{xy}(6,11) &= -\phi_{xy}(6,3) = -\gamma_2\frac{\delta a+1}{A^2}, \\
\end{aligned}
\begin{aligned}[c]
\phi_{xx}(6,2) &= \phi_{xx}(6,0) = -\gamma_1\left( \frac{\delta a-1}{B} \right)^2,\\
\phi_{yy}(6,2) &= \phi_{yy}(6,0) =  -\gamma_1\left( \frac{1}{B} \right)^2, \\
\phi_{xy}(6,2) &= -\phi_{xy}(6,0) = -\gamma_1\frac{\delta a-1}{B^2}. \\
\end{aligned}
\end{equation*}
Also,
\begin{equation}
\begin{split}
\phi_{xx}(6,5)  &= \phi_{xx}(6,9) = \phi_{yy}(6,7) = \phi_{yy}(6,10)= -\gamma_0, \\
\phi_{xx}(6,6) &= - \sum_{j}\left[ \phi_{xx}(6,j) \right] = 2\gamma_0 + 2\gamma_2\left( \frac{\delta a+1}{A} \right)^2 + 2\gamma_1\left( \frac{\delta a-1}{B} \right)^2, \\
\phi_{yy}(6,6)  &= -\sum_{j}\left[ \phi_{yy}(6,j) \right] = 2\gamma_0 + 2\gamma_2\left( \frac{1}{A} \right)^2 + 2\gamma_1\left( \frac{1}{B} \right)^2,
\end{split}
\end{equation}
Here, $j$ is summed over  nearest neighbors ($0,2,3,11$) and next-nearest neighbors ($5,7,9,10$).
All other force constants are zero.

\subsubsection{The dynamical $D$-matrix}
We are now ready to calculate the dynamical matrix ($D$-matrix) on the basis
\begin{align}
\{ u_{1,x}, u_{1,y}, u_{2,x}, u_{2,y} \}.
\end{align}
The dynamical $D$-matrix is defined from the equation of motion in the momentum space for the lattice vibrations,
\begin{align}
\label{eq:w2}
\omega^2 u_{n,\alpha}	= \sum_{m,\beta} {\cal D}_{\alpha\beta}(m,n) u_{m,\beta},
\end{align}
where $m,n\in\{1,2\}$ labels the two Sb atoms in one unit cell and the momentum label ${\bf q}=(q_x,q_y)$ is implicit in ${\cal D}_{\alpha\beta}(m,n) $.
The relation between the matrix elements of the $D$-matrix and the force constants of the preceding subsection is given by (cf. Fig.~\ref{sm_fig1}(a))
\begin{align}
{\cal D}_{\alpha\beta}(m,n) = \frac{1}{\sqrt{M_mM_n}} \sum_{i} \phi_{\alpha\beta} (m,i)  e^{-i{\bf q}\cdot {\bf R}_i},
\end{align}
where $M_n=M_m=M$ is the mass of Sb atoms. The precise meaning of the above summation is as follows:
\begin{align}
\begin{cases}
m=1,n=1: \; \sum_{i}\phi_{\alpha\beta}(m,i) e^{-i{\bf q}\cdot {\bf R}_i} \to \sum_{i=\{0,1,2,3,4\}} \phi_{\alpha\beta}(0,i) e^{-i{\bf q}\cdot {\bf R}_i},  \\
m=2,n=2: \; \sum_{i}\phi_{\alpha\beta}(m,i) e^{-i{\bf q}\cdot {\bf R}_i} \to \sum_{i=\{6,5,7,9,10\}} \phi_{\alpha\beta}(6,i) e^{-i{\bf q}\cdot {\bf R}_i}, \\
m=1,n=2: \; \sum_{i}\phi_{\alpha\beta}(m,i) e^{-i{\bf q}\cdot {\bf R}_i} \to \sum_{i=\{5,6,7,8\}}  \phi_{\alpha\beta}(0,i) e^{-i{\bf q}\cdot {\bf R}_i},  \\
m=2,n=1: \; \sum_{i}\phi_{\alpha\beta}(m,i) e^{-i{\bf q}\cdot {\bf R}_i} \to \sum_{i=\{0,2,3,11\}} \phi_{\alpha\beta}(6,i) e^{-i{\bf q}\cdot {\bf R}_i}.
\end{cases}
\end{align}
After a straightforward calculation, the elements of the dynamical $D$-matrix for the first Sb atom in one unit cell are given by
\begin{align}
\begin{split}
{\cal D}_{xx}(1,1) &= \frac{1}{M} \left\lbrack \phi_{xx}(0,0) + \phi_{xx}(0,1)e^{iq_xa}+\phi_{xx}(0,3)e^{-iq_xa} \right\rbrack \\
&= \frac{2}{M}\left\lbrack  \gamma_2\left( \frac{\delta a+1}{A} \right)^2  +  \gamma_1\left( \frac{\delta a-1}{B} \right)^2 +\gamma_0(1-\cos (q_xa)) \right\rbrack , \\
{\cal D}_{yy}(1,1) &= \frac{1}{M} \left\lbrack \phi_{yy}(0,0) + \phi_{yy}(0,2)e^{iq_ya}+\phi_{yy}(0,4)e^{-iq_ya} \right\rbrack \\
&= \frac{2}{M}\left\lbrack  \gamma_2\left( \frac{1}{A} \right)^2  +  \gamma_1\left( \frac{1}{B} \right)^2 +\gamma_0(1-\cos (q_ya)) \right\rbrack , \\
{\cal D}_{xy}(1,1) &= {\cal D}_{yx}(1,1) = 0.
\end{split}
\end{align}
For the second Sb atom in one unit cell, we have
\begin{align}
{\cal D}_{xx}(2,2)={\cal D}_{xx}(1,1), {\cal D}_{yy}(2,2)={\cal D}_{yy}(1,1), \text{ and } {\cal D}_{xy}(2,2)={\cal D}_{yx}(2,2)=0.
\end{align}
Concerning the off-diagonal terms between these two Sb atoms, we have
\begin{align}
{\cal D}_{xx}(1,2) &= \frac{1}{M} \left\lbrack  \phi_{xx}(0,5)e^{i\frac{a}{2}\left( q_x(\delta a+1)+q_y\right)} + \phi_{xx}(0,8)e^{i\frac{a}{2}\left( q_x(\delta a+1)-q_y\right)} + \phi_{xx}(0,6)e^{i\frac{a}{2}\left( q_x(\delta a-1)+q_y\right)} +\phi_{xx}(0,7)e^{i\frac{a}{2}\left( q_x(\delta a-1)-q_y\right)} \right\rbrack, \nonumber \\
&= \frac{2}{M}\cos\left(q_y\frac{a}{2}\right) \left\lbrack - \gamma_2\left( \frac{\delta a+1}{A} \right)^2 e^{iq_x\frac{a}{2}(\delta a+1)} -  \gamma_1\left( \frac{\delta a-1}{B} \right)^2 e^{iq_x \frac{a}{2}(\delta a-1) }   \right\rbrack, \nonumber \\
{\cal D}_{yy}(1,2) &= \frac{2}{M}\cos\left(q_y\frac{a}{2}\right) \left\lbrack - \gamma_2\left( \frac{1}{A} \right)^2 e^{iq_x\frac{a}{2}(\delta a+1)} -  \gamma_1\left( \frac{1}{B} \right)^2 e^{iq_x \frac{a}{2}(\delta a-1) }   \right\rbrack, \\
{\cal D}_{xy}(1,2) &= i\frac{2}{M}\sin\left( q_y\frac{a}{2} \right) \left\lbrack -\gamma_2\frac{\delta a+1}{A^2} e^{iq_x\frac{a}{2}(\delta a+1)} -\gamma_1\frac{\delta a-1}{B^2} e^{iq_x\frac{a}{2}(\delta a-1)}  \right\rbrack, \nonumber \\
{\cal D}_{yx}(1,2) &={\cal D}_{xy}(1,2). \nonumber
\end{align}
Likewise, $\mathcal{D}_{\tau\tau'}(2,1)=\mathcal{D}_{\tau\tau'}^\ast(1,2)$.
Therefore, the four-by-four $D$-matrix is given by
\begin{align}
\mathcal{D} = \left\lbrack\begin{array}{cccc}
{\cal D}_{xx}(1,1) & 0 & {\cal D}_{xx}(1,2) & {\cal D}_{xy}(1,2) \\
0 & {\cal D}_{yy}(1,1) & {\cal D}_{xy}(1,2) & {\cal D}_{yy}(1,2) \\
{\cal D}_{xx}^\ast(1,2) &  {\cal D}_{xy}^\ast(1,2)  & {\cal D}_{xx}(2,2) & 0 \\
{\cal D}_{xy}^\ast(1,2) &  {\cal D}_{yy}^\ast(1,2) & 0 & {\cal D}_{yy}(2,2)
\end{array} \right\rbrack.
\end{align}
The vibration spectrum is the solution of $\vert \mathcal{D}({\bf q}) - \omega^2 \vert=0$.
To gain analytical understanding, hereafter we consider the vibrational modes at the $\Gamma$ point ($q_x=q_y=0$), which has the full lattice symmetry.
Then, the $D$-matrix is
\begin{align}\label{sm-eq-d-matrix1}
\mathcal{D}({\bf q}=0) = \left\lbrack  \begin{array}{cccc}
D_1 & 0 & D_3 & 0 \\
0 & D_2 & 0 & D_4 \\
D_3 & 0 & D_1 & 0 \\
0 & D_4 & 0 & D_2
\end{array}
\right\rbrack,
\end{align}
where $D_1= \frac{2}{M}\left\lbrack \gamma_2\left( \frac{\delta a+1}{A} \right)^2 + \gamma_1\left( \frac{\delta a-1}{B} \right)^2  \right\rbrack $, $D_2= \frac{2}{M}\left\lbrack \gamma_2\left( \frac{1}{A} \right)^2 + \gamma_1\left( \frac{1}{B} \right)^2  \right\rbrack $, $D_3=-D_1$, and $D_4=-D_2$.
By diagonalizing ${\cal D}({\bf 0})$, we obtain the frequencies and displacement vectors for the four in-plane vibrational modes:
\begin{align}
\begin{split}
\omega_{a,A_1} &=0 \text{ and } \vec{u}_{a,A_1} = (1,0,1,0)^T/\sqrt{2},   \\
\omega_{a,B_1} &=0 \text{ and }   \vec{u}_{a,B_1} = (0,1,0,1)^T/\sqrt{2}, \\
\omega_{o,A_1} &=2D_1 \text{ and } \vec{u}_{o,A_1} = (-1,0,1,0)^T/\sqrt{2},  \\
\omega_{o,B_1} &=2D_2 \text{ and } \vec{u}_{o,B_1} = (0,-1,0,1)^T/\sqrt{2}.
\end{split}
\end{align}
For acoustic phonons, ${\bf q}=0$ corresponds to a rigid translation of the crystal (hence $\omega=0$) and its phonon modes are shown in Fig.~\ref{sm_fig1}(b).
For optical phonons, the two Sb atoms in one unit cell move $180^\circ$ out of phase, shown in Fig.~\ref{sm_fig1}(c).
As discussed next, this is consistent with the symmetry analysis.

\subsection{Symmetry construction of the dynamical matrix at $q=0$}
In the preceding subsection, we have constructed the $D$-matrix by using the force constant approach with harmonic approximation.
In this subsection, we recover the same for of the $D$-matrix using symmetry arguments.
We begin by rewriting the general form of dynamical $D$-matrix  in the $\{ u_{1,x}, u_{1,y}, u_{2,x}, u_{2,y} \}$ basis as
\begin{align}
{\cal  D} = \left(\begin{array}{cccc}
{\cal D}_{xx}(1,1) & {\cal D}_{x y} (1,1) & {\cal D}_{xx}(1,2) & {\cal D}_{xy}(1,2) \\
{\cal D}_{yx}(1,1) & {\cal D}_{yy}(1,1) & {\cal D}_{yx} (1,2) & {\cal D}_{yy}(1,2) \\
{\cal D}_{xx}(2,1) & {\cal D}_{xy}(2,1) & {\cal D}_{xx}(2,2) & {\cal D}_{xy}(2,2) \\
{\cal D}_{yx}(2,1) & {\cal D}_{yy}(2,1) & {\cal D}_{yx}(2,2) & {\cal D}_{yy}(2,2)
\end{array} \right).
\end{align}
Let us consider the symmetry constraints due to the symmetry operators $\{C_{2x},\sigma_v(xz)\}$, following Ref.~[\onlinecite{Maradudin1968}].
The transformation of the dynamical matrix under crystal symmetry operations is given by
\begin{equation}
\label{eq:Dsym}
\boldsymbol{\Gamma}({\bf q};\{ R|{\bf v}(R)+{\bf x}(l)\}) {\cal D}({\bf q}) \boldsymbol{\Gamma}^{-1}({\bf q};\{ R|{\bf v}(R)+{\bf x}(l) \})  = {\cal D}(R {\bf q}),
\end{equation}
where the representation matrix $\boldsymbol{\Gamma}$ of the symmetry operator $\{ R|{\bf v}(R)+{\bf x}(l) \}$ is given by
\begin{equation}
\Gamma_{\alpha\beta}(m,n|{\bf q};\{ R|{\bf v}(R)+{\bf x}\}) = R_{\alpha\beta}\delta(m,F_o(n;R)) e^{i{\bf q}\cdot[\{ R|{\bf v}(R)+{\bf x}(l)\}^{-1}{\bf x}(m)-{\bf x}(n)]}.
\end{equation}
Here, $m, n\in \{{1,2}\}$ denote the two Sb atoms in a unit cell, $\alpha,\beta\in\{x,y\}$, $R$ is a real $2\times 2$ orthogonal matrix representation of one of the proper or improper rotations of the point group, ${\bf v}(R)$ is a vector which is smaller than any primitive translation vector of the crystal (in symmorphic space groups such as ours, ${\bf v}(R)={\bf 0}$ for all $R$), ${\bf x}(l)$ is the position of the origin of $l-$th unit cell,  ${\bf x}(m)$ is the position of an atom $m$ relative to the origin of the unit cell, and $\delta(m,F_o(n;R))$ indicates which atom $m$ is brought into the position of $n$ under the symmetry operation.

Hereafter, we focus on $R\in\{C_{2x}, \sigma_v(xz)\}$ and ${\bf q}=0$. Thus,
\begin{align}
\Gamma_{\alpha\beta}(m, n|{\bf 0};\{ R|{\bf v}(R)+{\bf x}(l)\}) = R_{\alpha\beta} \delta(m,n),
\end{align}
where $R=\text{Diag}[1,-1]$.
Therefore, $\boldsymbol{\Gamma}=\text{Diag}[1,-1,1,-1]$ and, according to Eq.~(\ref{eq:Dsym}), we arrive at
\begin{align}
{\cal D}= \left(\begin{array}{cccc}
{\cal D}_{xx}(1,1) & 0 & {\cal D}_{xx}(1,2) & 0 \\
0 & {\cal D}_{yy}(1,1) & 0 & {\cal D}_{yy}(1,2) \\
{\cal D}_{xx}(2,1) & 0 & {\cal D}_{xx}(2,2) & 0 \\
0 & {\cal D}_{yy}(2,1) & 0 & {\cal D}_{yy}(2,2)
\end{array} \right).
\end{align}
The dynamical matrix must be Hermitian, ${\cal D}^\dagger({\bf q})={\cal D}({\bf q})$. In addition, time-reversal symmetry requires ${\cal D}({\bf q})={\cal D}^\ast(-{\bf q})$.
As a result, the $D$-matrix becomes
\begin{align}
{\cal D}= \left(\begin{array}{cccc}
{\cal D}_{xx}(1,1) & 0 & {\cal D}_{x x}(1,2) & 0 \\
0 & {\cal D}_{yy}(1,1) & 0 &{\cal D}_{yy}(1,2) \\
{\cal D}_{xx}(1,2) & 0 & {\cal D}_{xx}(2,2) & 0 \\
0 & {\cal D}_{yy}(1,2) & 0 & {\cal D}_{yy}(2,2)
\end{array} \right),
\end{align}
where all the matrix elements are real.
One more constraint on ${\cal D}$ comes from the fact that the lattice energy does not change for an overall translation of the whole lattice, i.e. the acoustic phonon should have zero energy for ${\bf q}=0$.
The overall translation of the whole lattice along the $x$-direction is described by the vector $(u_{1,x}, u_{1,y}, u_{2,x}, u_{2,y})= (1, 0, 1, 0)$. Thus, we expect
\eq{
	{\cal D}(1,0,1,0)^T=0.
}
Similarly, the overall translation of the whole lattice along the $y$ direction gives rise to
\eq{
	{\cal D} (0,1,0,1)^T=0.
}
These two equations result in
\begin{align}
{\cal D}_{xx}(1,1)={\cal D}_{xx}(2,2)=-{\cal D}_{xx}(1,2), \text{ and } {\cal D}_{yy}(1,1)={\cal D}_{yy}(2,2)=-{\cal D}_{yy}(1,2),
\end{align}
which makes the form of the $D$-matrix identical to the one obtained above from the force constant model (see Eq.~\eqref{sm-eq-d-matrix1}).

\section{Electron-phonon interaction Hamiltonian for optical phonons}
In this section, we derive the low-energy effective Hamiltonian for the interaction between electrons and optical phonons in BaMnSb$_2$. The main objective is to obtain Eq.~(5) of the main text.
We begin by briefly reviewing key aspects of low-energy electrons~[\onlinecite{liu2019}],  which are massive Dirac fermions located at ${\bf K}_\pm = (\pi/a,\pm k_{y_0} )$:
\begin{itemize}
	\item[1.] Only the $p_x$ and $p_y$ orbitals of one atom in a unit cell contribute to the states near the Fermi energy. Thus, the basis of electronic states is given by
	\begin{align}\label{sm-eq-c-basis}
	C_{\mathbf{k}}^\dagger = \left( c_{p_x,\uparrow}^\dagger, c_{p_x,\downarrow}^\dagger, c_{p_y,\uparrow}^\dagger, c_{p_y,\downarrow}^\dagger \right).
	\end{align}
	\item[2.] The transformation of the electronic states under symmetry operations is described by
	\begin{align}\label{sm-eq-symmetry-on-c}
	\begin{split}
	C_{2x} C_{\mathbf{K}_+}^\dagger C_{2x}^{-1} &= C_{C_{2x}\mathbf{K}_+}^\dagger (-i\tau_3\sigma_1),\\
	\mathcal{T} C_{\mathbf{K}_+}^\dagger \mathcal{T}^{-1} &= C_{-\mathbf{K}_+}^\dagger(i\tau_0\sigma_2),\\
	\mathcal{P} C_{\mathbf{K}_+}^\dagger \mathcal{P}^{-1} &= C_{\mathbf{K}_+}^\dagger (i\tau_3\sigma_3	),
	\end{split}
	\end{align}
	where $\tau_i$ and $\sigma_i$ are Pauli matrices acting in orbital and spin subspaces, respectively.
	Also, $\mathcal{T}$ is the time-reversal operator. Both $C_{2x}$ and $\mathcal{T}$ interchange the valleys ${\bf K_+}$ and ${\bf K_-}$. Thus, on each valley, only the combined symmetry $\mathcal{P}=C_{2x}\mathcal{T}$ remains.
\end{itemize}
We now derive the electron-phonon (e-ph) Hamiltonian following Ref.~[\onlinecite{Saha2015},\onlinecite{Rinkel2017}].
The displacement of atoms away from their equilibrium positions gives rise to a deformation potential $\delta U$, which couples to the electron density as
\begin{align}
\mathcal{H}_{e-ph} &= \int d^2 r \, \Psi^\dagger(\mathbf{r}) \Psi(\mathbf{r})\delta U(\mathbf{r}), \\
\delta U(\mathbf{r}) &= \sum_{{\bf l},s} \left[ U(\mathbf{r}-\mathbf{R}_{\mathbf{l},s} - \mathbf{Q}_{\mathbf{l},s}) - U(\mathbf{r}-\mathbf{R}_{\mathbf{l},s}) \right]
\simeq \sum_{{\bf l},s} \mathbf{Q}_{\mathbf{l},s} \cdot \frac{\partial U(\mathbf{r}-\mathbf{R}_{\mathbf{l},s})}{\partial \mathbf{R}_{\mathbf{l},s}},
\end{align}
where  $U(\mathbf{r})=\sum_{\mathbf{l},s} U_s(\mathbf{r}-\mathbf{R}_{\mathbf{l},s})$ is the periodic lattice potential,
${\bf l}$ is the position of the $l-$th unit cell, $s$ labels atoms inside the unit cell,
$\mathbf{R}_{\mathbf{l},s}=\mathbf{l}+\mathbf{r}_s$ is the position of atom $s$ in unit cell $l$,
\begin{align}
\mathbf{Q}_{\mathbf{l},s} = \sum_{\mathbf{q},\lambda} \sqrt{\frac{\hbar}{2M_sN\omega_{\mathbf{q},\lambda}}} e^{i\mathbf{q}\cdot(\mathbf{l}+\mathbf{r}_s)} \mathbf{P}_{{\bf q},s}^\lambda \left(b_{\mathbf{q},\lambda}+b^\dagger_{-\mathbf{q},\lambda}\right)
\end{align}
is the displacement operator for atom $s$ in unit cell $l$, $b_{\mathbf{q},\lambda}$ is an operator that annihilates a phonon mode $\lambda$ with momentum $\mathbf{q}$, $\omega_{\lambda}(\mathbf{q})$ is the phonon frequency, $N$ is the number of unit cells in the crystal, $M_s$ is
the mass of atom $s$ and ${\bf P}_{{\bf q}, s}^\lambda$ is the polarization vector corresponding to atom $s$ in phonon mode $\lambda$ with momentum $\mathbf{q}$.
In addition, the fermion field operator $\Psi(\mathbf{r})$ can be expanded into a sum over Bloch wave functions, $\Psi(\mathbf{r})=\frac{1}{\sqrt{\mathcal{A}}}\sum_{\mathbf{k}}\sum_{\tau\sigma, {\bf k}}e^{i\mathbf{k}\cdot\mathbf{r}} u_{\tau\sigma}(\mathbf{r}) c_{\sigma\tau}(\mathbf{k}) $, where $\mathcal{A}$ is the area of the 2D sample.
Here, $u_{\tau\sigma, {\bf k}}(\mathbf{r})$ is the periodic part of the Bloch wave function, which satisfies $u_{\tau\sigma, {\bf k}}(\mathbf{r})=u_{\tau\sigma, {\bf k}}(\mathbf{r}+\mathbf{l})$ and is calculated by assuming that the atoms are in their equilibrium positions.

The low-energy electrons are contained in the vicinity of two valleys $K_\pm$, which are well-separated in momentum space. Then, we truncate the momentum sum in the fermion field operator $\Psi(\mathbf{r})$ around the vicinity of the two valleys,
\begin{align}
\Psi(\mathbf{r})=\frac{1}{\sqrt{\mathcal{A}}}\sum_{|\mathbf{k}|<\Lambda} \sum_{\kappa} e^{i\mathbf{K}_\kappa\cdot\mathbf{r}} \sum_{\tau\sigma} e^{i\mathbf{k}\cdot\mathbf{r}} u_{\tau\sigma,\kappa}(\mathbf{r}) c_{\sigma\tau,\kappa}(\mathbf{k}) + \mathcal{O}\left(\frac{k}{\Lambda}\right),
\end{align}
where $\kappa=\pm$ label the two valleys, now ${\bf k}$ is the momentum measured from a valley and $\Lambda$ is a cutoff under which the linear dispersion of Dirac electrons is valid.
Based on this approximation, we can derive the e-ph coupling Hamiltonian at the two valleys separately.

Neglecting intervalley electron-phonon scattering and performing some straightforward manipulations (details can be found in Ref.~[\onlinecite{Saha2015},\onlinecite{Rinkel2017}]), the e-ph Hamiltonian becomes
\begin{align}
\label{sm-eq-eph-ham-optical}
\mathcal{H}_{e-ph} &= \sum_{{\bf k},{\bf q}}\sum_{\lambda,\kappa}\sum_{\sigma\tau,\sigma'\tau'} g^{\lambda,\kappa}_{\sigma\tau,\sigma'\tau'}(\mathbf{q}) Q_\lambda(\mathbf{q}) c_{\sigma\tau,\kappa}^\dagger ({\bf k}) c_{\sigma'\tau',\kappa}(\mathbf{k}-\mathbf{q}),\\
g^{\lambda,\kappa}_{\sigma\tau,\sigma'\tau'}(\mathbf{q}) &= \frac{1}{\sqrt{N}\mathcal{A}} \sum_{s} \int d^2 r\, u_{\sigma\tau,\kappa}^\ast(\mathbf{r}) u_{\sigma' \tau',\kappa}(\mathbf{r}) e^{-i\mathbf{q}\cdot\mathbf{r}}
\times\left\lbrack \mathbf{P}^\lambda_{{\bf q},s} \cdot \frac{\partial U(\mathbf{r}-\mathbf{r}_s)}{\partial \mathbf{r}_s} \right\rbrack,	
\label{sm-eq-g-optical}
\end{align}
where $\lambda=\{A_1,B_1\}$ labels the two optical in-plane phonon modes and $Q_\lambda(\mathbf{q}) = \sqrt{\frac{\hbar}{2M \omega_{\lambda}(\mathbf{q})}} \left(b_{\mathbf{q},\lambda}+b^\dagger_{-\mathbf{q},\lambda}\right)$ is the displacement operator for mode $\lambda$ at momentum ${\bf q}$.
Note that $\hbar=1$ is assumed in the main text.
Below,
we only consider the e-ph coupling for long-wavelength phonons.
Moreover, the Hermiticity of the e-ph coupling Hamiltonian requires that $\mathbf{g}^{\lambda,\kappa}(-\mathbf{q}) = \left[\mathbf{g}^{\lambda,\kappa}(\mathbf{q})\right]^\dagger$, where $\mathbf{g}^{\lambda,\kappa}({\bf q})$ is a $4\times 4$ matrix whose matrix elements are $g^{\lambda,\kappa}_{\sigma\tau,\sigma'\tau'}(\mathbf{q})$.

\subsection{Symmetry analysis for the electron-phonon coupling vertex}
In this subsection, we focus on the symmetry analysis for the e-ph coupling Hamiltonian in Eq.~\eqref{sm-eq-eph-ham-optical}.
For notational simplicity, we relabel the two atoms in a unit cell as Sb$_{1}$ and Sb$_2$ (this is not to be confused with the labels in Fig. \ref{sm_fig1}(a)).
We restrict ourselves to long-wavelength optical phonons, for which the e-ph coupling matrix $\mathbf{g}^{\lambda,\kappa}(\mathbf{q})$ is independent of $\mathbf{q}$ to leading order in ${\bf q}$.
This approximation relies on $\mathbf{g}^{\lambda,\kappa}({\bf q})$ being an analytic function of ${\bf q}$ at $q\to 0$. Analyticity is expected in our problem, due to the non-degeneracy of low-energy electronic and phononic states.
Moreover, the matrix elements of $\mathbf{g}^{\lambda,\kappa}({\bf 0})$ are real due to the Hermiticity condition.

In the low-energy effective theory for electrons, the spin degree of freedom is locked to the valley degree of freedom.
As a result, the e-ph coupling for long-wavelength phonons is spin-conserving.
We will furthermore assume that the e-ph coupling is spin-independent.
Then,
we drop the spin index and work with the spinless basis $ C_{\mathbf{k}}^\dagger = ( c_{p_x}^\dagger, c_{p_y}^\dagger )$.
At the same time, the four-by-four matrix $\mathbf{g}^{\lambda,\kappa}({\bf 0})$ is reduced to a two-by-two matrix, with elements given by
\begin{align}\label{sm-eq-g-matrix-2-by-2}
g^{\lambda,\kappa}_{\tau,\tau'}(0) = \frac{1}{\sqrt{N}\mathcal{A}} \sum_{s} \int d^2\mathbf{r} \; u_{\tau,\kappa}^\ast(\mathbf{r}) u_{\tau',\kappa}(\mathbf{r})
\times\left\lbrack \mathbf{P}^\lambda_{{\bf q},s} \cdot \frac{\partial U(\mathbf{r}-\mathbf{r}_s)}{\partial \mathbf{r}_s} \right\rbrack.
\end{align}

Next, we determine the matrix structure of $\mathbf{g}^{\lambda,\kappa}({\bf 0})$ for $A_1$ and $B_1$ in-plane optical phonons, using symmetry arguments.
For brevity, we hereafter drop the ${\bf q}=0$ indicator from $\mathbf{g}^{\lambda,\kappa}$.
We start with the e-ph Hamiltonian for electrons at the $K_+$ valley,
\begin{align}
\mathcal{H}_{e-ph}^+ = C_{\mathbf{K_+}}^\dagger \left[\mathbf{g}^{\lambda,+} Q_\lambda(\mathbf{q}\to0) \right] C_{\mathbf{K_+}}.
\end{align}
The invariance of $\mathcal{H}_{e-ph}^+$ under the combined symmetry operator $\mathcal{P}=C_{2x}\mathcal{T}$ implies $ \mathcal{P} \mathcal{H}_{e-ph}^+\mathcal{P}^{-1} ={\cal H}_{e-ph}^+$, where
\begin{align}\label{sm-eq-c2xT-e-ph-matrix}
\mathcal{P} \mathcal{H}_{e-ph}^+\mathcal{P}^{-1} = C_{\mathbf{K_+}}^\dagger \left[ (i\tau_3) (x_\lambda \mathbf{g}^{\lambda,+} Q_\lambda(\mathbf{q}\to0))^\ast (-i\tau_3) \right] C_{\mathbf{K_+}}
\end{align}
and the coefficient $x_\lambda$ (which will be either $+1$ or $-1$, depending on $\lambda$) is defined via $\mathcal{P} Q_\lambda \mathcal{P}^{-1}=x_\lambda Q_\lambda$.
Therefore, we have $ \tau_3 (\mathbf{g}^{\lambda,+})^\ast \tau_3 = x_\lambda \mathbf{g}^{\lambda,+}$.

For the $\lambda=A_1$ phonon mode, the Sb$_1$ atom moves along the $x$ direction. Hence, $x_{A_1}=+1$.
We do not need to consider the Sb$_2$ atom in the unit cell, since there are no electron states of Sb$_2$ near the Fermi surface.
Therefore, we have $ \tau_3 (\mathbf{g}^{A_1,+})^\ast \tau_3 = \mathbf{g}^{A_1,+}$, which leads to
\begin{align}
\mathbf{g}^{A_1,+} = g_0^o\tau_0 + g_2^o\tau_2 + g_3^o\tau_3.
\end{align}
In the $K_-$ valley, time-reversal symmetry imposes
\begin{equation}
\mathbf{g}^{{A_1},-} = g_0^o\tau_0 - g_2^o\tau_2 + g_3^o\tau_3.
\end{equation}

For $\lambda=B_1$, the Sb$_1$ atom moves along the $y$ direction.
Then, $x_{B_1}=-1$ and $ \tau_3 (\mathbf{g}^{B_1,+})^\ast \tau_3 = -\mathbf{g}^{B_1,+}$, which leads to
\begin{equation}
\mathbf{g}^{B_1,+} = g_1^o\tau_1.
\end{equation}
For the $K_-$ valley, time-reversal symmetry imposes
\begin{equation}
\mathbf{g}^{B_1,-} = g_1^o\tau_1.
\end{equation}
In summary,
\begin{align}
\begin{split}
\mathbf{g}^{o,A_1,s} &= g_0^o \tau_0 + s g_2^o \tau_2 + g_3^o\tau_3, \\
\mathbf{g}^{o,B_1,s} &= g_1^o\tau_1,
\end{split}
\end{align}
where $g_{0,1,2,3}^o$ are material-dependent real parameters.
This coincides with Eq.~(5) of the main text.

\subsection{Analysis by deformation potential theory}
In this subsection, we present an alternative derivation of the e-ph $\mathbf{g}^{\lambda,\kappa}$ matrix.
The starting point is Eq.~\eqref{sm-eq-g-matrix-2-by-2}.
Again, all the low-energy electronic states are coming from the Sb$_1$ atom, and the e-ph coupling is assumed to be independent of spin.
Therefore, the expression for the e-ph coupling at the $K_+$ valley is given by
\begin{align}
g^{\lambda,+}_{\tau,\tau'}(\mathbf{q}) &= \frac{1}{\sqrt{N}\mathcal{A}} \int d^2 r \, u_{\tau,+}^\ast(\mathbf{r}) u_{\tau',+}(\mathbf{r}) e^{-i\mathbf{q}\cdot\mathbf{r}}
\times\left\lbrack \mathbf{P}^\lambda_{\bf q} \cdot \frac{\partial U(\mathbf{r}-\mathbf{r}_1)}{\partial \mathbf{r}_1} \right\rbrack,	
\end{align}
where $\mathbf{r}=(x,y)$, $\mathbf{r}_1=(x_1,y_1)$, $\mathbf{P}^A_{\bf q}=(1,0)$ and $\mathbf{P}^B_{\bf q}=(0,1)$ (when $q\to 0$).
Then, we study the constraints due to the combined symmetry operator $\mathcal{P}$.
Firstly, it gives rise to $\mathcal{P}: \, (x,y)\to(x,-y)$.
And the Bloch wave basis are transfered as $\mathcal{P}:\, u_{p_x}(x,y) \to u_{p_x}^\ast(x,-y) $ and $u_{p_y}(x,y) \to -u_{p_y}^\ast(x,-y) $.
As for the lattice potential, $\mathcal{P}:\, U(x-x_1,y-y_1) \to U(x-x_1,-y+y_1)$.
Therefore,
\begin{itemize}
	\item[(1)] For the $\lambda=A_1$ phonon, all the element of $g^{A_1,+}$ matrix in the orbital subspace,
	\begin{align}
	g^{A_1,+}_{\tau,\tau'} = \frac{1}{\sqrt{N}\mathcal{A}} \int d^2 r\,\frac{\partial U(\mathbf{r}-\mathbf{r}_1)}{\partial x_1} u_{\tau,+}^\ast(x,y) u_{\tau',+}(x,y),
	\end{align}
	where $\mathbf{r}_1=(x_1,0)$ because all the Sb$_1$ atoms move along the $x$-direction.
	Under the combined $\mathcal{P}$ symmetry operator, we have
	\begin{align}
	\begin{split}
	g^{A_1,+}_{p_x,p_x} &= \frac{1}{\sqrt{N}\mathcal{A}} \int d^2r\, \frac{\partial U(\mathbf{r}-\mathbf{r}_1)}{\partial x_1} u_{p_x,+}^\ast(x,y) u_{p_x,+}(x,y),\\
	&\to \frac{1}{\sqrt{N}\mathcal{A}} \int d^2 r\, \frac{\partial U(x-x_1,-y)}{\partial x_1} u_{p_x,+}(x,-y) u_{p_x,+}^\ast(x,-y) = g^{A_1,+}_{p_x,p_x},
	\end{split}
	\end{align}
	where we apply $P$ to everything inside the integral of the first line, and we replace $y\to-y$ for the second line.
	Similarly, one can easily check that
	\begin{align}
	g^{A_1,+}_{p_y,p_y} \to g^{A_1,+}_{p_y,p_y} , \quad  g^{A_1,+}_{p_x,p_y} \to -  g^{A_1,+}_{p_y,p_x} .
	\end{align}
	
	From the above results, we have
	\begin{align}
	\left(\begin{array}{cc}
	g^{A_1,+}_{p_x,p_x} 	& g^{A_1,+}_{p_x,p_y} \\
	g^{A_1,+}_{p_y,p_x}  	& g^{A_1,+}_{p_y,p_y}
	\end{array} \right) =  \left(\begin{array}{cc}
	g^{A_1,+}_{p_x,p_x} 	& -g^{A_1,+}_{p_y,p_x} \\
	-g^{A_1,+}_{p_x,p_y}	& g^{A_1,+}_{p_y,p_y}
	\end{array} \right) \Rightarrow  \left(\begin{array}{cc}
	g^{A_1,+}_{p_x,p_x} 	& g^{A_1,+}_{p_x,p_y} \\
	g^{A_1,+}_{p_y,p_x}	    & g^{A_1,+}_{p_y,p_y}
	\end{array} \right)= g_0^o \tau_0 + g_2^o \tau_2 + g_3^o\tau_3.
	\end{align}
	For the $K_-$ valley, the time-reversal symmetry leads to $\mathbf{g}^{{A_1},-} = g_0^o\tau_0 - g_2^o\tau_2 + g_3^o\tau_3$.
	\item[(2)]  For the $\lambda=B_1$ phonon, the matrix elements of  the $\mathbf{g}^{B_1,+}$ matrix in the orbital subspace read
	\begin{align}
	g^{B_1,+}_{\tau,\tau'} = \frac{1}{\sqrt{N}\mathcal{A}} \int d^2\mathbf{r} \frac{\partial U(\mathbf{r}-\mathbf{r}_1)}{\partial y_1} u_{\tau,+}^\ast(\mathbf{r}) u_{\tau',+}(\mathbf{r}).
	\end{align}
	Please note that $\frac{\partial U(\mathbf{r}-\mathbf{r}_1)}{\partial y_1}$ also changes sign under the $\mathcal{P}$ symmetry operator.
	In the same way, we have
	\begin{align}
	g^{B_1,+}_{p_x,p_x} \to -g^{B_1,+}_{p_x,p_x}, \quad
	g^{B_1,+}_{p_y,p_y} \to -g^{B_1,+}_{p_y,p_y}, \quad
	g^{B_1,+}_{p_x,p_y} \to  g^{B_1,+}_{p_y,p_x}.
	\end{align}
	Similar relations apply for the $K_-$ valley. Then, $\mathbf{g}^{B_1,+}=\mathbf{g}^{B_1,-} = g_1^o\tau_1$.
\end{itemize}
In sum, the two methods of this and the preceding subsection give consistent results to construct the e-ph Hamiltonian.
As a result, the full Hamiltonian can be written as follows,
\begin{align}
\label{sm-eq-h0-electron-2-by-2}
&\text{Electron Hamiltonian: } h_\pm(\mathbf{k}) = \pm v_0(k_y\tau_3 + k_x\tau_1) \pm m_0 \tau_2,\\
\label{sm-eq-h0-electron-2-by-2-p1}
&\text{e-ph Hamiltonian at $K_+$: } {\cal H}^{+}_{e-ph} = \sum_{\mathbf{k},\mathbf{q}} C^\dagger_\mathbf{k} \left\lbrack \left( g_0^o \tau_0 + g_2^o \tau_2 + g_3^o\tau_3 \right)Q^A_\mathbf{q} +  g_1^o\tau_1 Q^B_\mathbf{q} \right\rbrack C_{\mathbf{k}+\mathbf{q}}    , \\
\label{sm-eq-h0-electron-2-by-2-p2}
&\text{e-ph Hamiltonian at $K_-$: } {\cal H}^{-}_{e-ph} = \sum_{\mathbf{k},\mathbf{q}} C^\dagger_\mathbf{k} \left\lbrack \left( g_0^o \tau_0 - g_2^o \tau_2 + g_3^o\tau_3 \right)Q^A_\mathbf{q} +  g_1^o\tau_1 Q^B_\mathbf{q} \right\rbrack C_{\mathbf{k}+\mathbf{q}},
\end{align}
where ${\bf k}$ is the momentum measured from a valley and the $g$-parameters depend on the material details.

	\subsection{Phonon-induced pseudo-gauge fields for Dirac electrons}
	In this section, we show that the e-ph coupling leads to electronic pseudo-gauge fields. 
	This can be seen by rewriting the relevant Hamiltonians in real space.
	\begin{align}
	H &=\sum_s \int d^2\mathbf{r} \, c^\dagger_s(\mathbf{r}) \left\{ s v_0\left\lbrack (-i \partial_y+A_y)\tau_3 + (-i \partial_x+A_x)\tau_1 \right\rbrack + s m_0 \tau_2 + A_0 \tau_0 \right\} c_s(\mathbf{r}) , \\
	H_{e-ph} &= \sum_s\int d^2\mathbf{r} \, c^\dagger_s(\mathbf{r}) \left\{  \left( g_0^o \tau_0 + s g_2^o \tau_2 + g_3^o\tau_3 \right)Q_{A_1}(\mathbf{r}) +  g_1^o\tau_1 Q_{B_1}(\mathbf{r})  \right\} c_s(\mathbf{r})\ ,
	\end{align}
	where we have included the $U(1)$ gauge field $A_\mu$, the matrix is $(-,+,+)$, the unit system is chosen such that $\hbar=c=1$, we have absorbed the elementary charge $e$ into $A_\mu$, and we have used 
	\eqa{\label{eq:FT}
		& c^{\dagger}_{s,\bsl{k}}= \frac{1}{\sqrt{\mathcal{A}}}\int d^2 r e^{\ii \bsl{k}\cdot\bsl{r}}c^{\dagger}_s(\mathbf{r}) \\
		& Q_{\lambda,\bsl{k}}=\frac{1}{\mathcal{A}} \int d^2 r e^{\ii \bsl{k}\cdot\bsl{r}}Q_{\lambda}(\mathbf{r})  .
	}
	Here the unit for the above operators are: $[c^{\dagger}_s(\mathbf{r})]  = [nm^{-1}] = [meV]$,  $ [c^{\dagger}_{s,\bsl{k}}] = [1]$, $[Q_{\lambda,\bsl{k}}]=[nm]=[meV^{-1}]$, and $[Q_{\lambda}(\mathbf{r})] =[nm]=[meV^{-1}]$.
	Notice that $[nm]=[meV^{-1}]$ in the case of $\hbar=c=1$.
	Adding $H_{e-ph}$ to $H$ gives
	\begin{align}\label{eq:pseudo-gauge}
	H_{tot} = \sum_s \int d^2\mathbf{r} \, c^\dagger_s(\mathbf{r}) \left\{  \widetilde{A}_{s,0} \tau_0 + s v_0\left\lbrack (-i\partial_y + \widetilde{A}_{s,y})\tau_3 + (-i\partial_x+\widetilde{A}_{s,x})\tau_1 \right\rbrack + s (m_0+g_2^o Q_{A_1}(\mathbf{r})) \tau_2 \right\} c_s(\mathbf{r})\ ,
	\end{align}
	where $\widetilde{A}_{s,0}=A_0 + g_0^o Q_{A_1}(\mathbf{r})$, $\widetilde{A}_{s,x}=A_x +  s\frac{1}{v_0} g_1^oQ_{B_1}(\mathbf{r})$, and $\widetilde{A}_{s,y}=A_y +  s\frac{1}{v_0} g_3^oQ_{A_1}(\mathbf{r})$.
	As a result, we can define the pseudo-gauge field at $K_s$ as
	\begin{align}\label{eq:A_pse}
	{ A}_{pse,\mu}^s(\mathbf{r}) = \left( g_0^o Q_{A_1}(\mathbf{r}), s\frac{1}{v_0} g_1^oQ_{B_1}(\mathbf{r}), s\frac{1}{v_0} g_3^oQ_{A_1}(\mathbf{r})  \right) _\mu,
	\end{align}
	where $\mu=0,1,2$ and the pseudo-gauge field couples to the electrons in the same way as the $U(1)$ gauge field at each valley.
	The unit of ${ A}_{pse,\mu}^s(\mathbf{r}) $ is [meV], since $[v_0]=[meV\cdot nm] = [1]$ and $[g_{0,1,3}^o]=[meV\cdot nm^{-1}]=[meV^2]$.
	In Eq.\,(4) of the main text, we show the above pseudo-gauge field in the momentum space.

\section{Electron-phonon interaction Hamiltonian for acoustic phonons}
In this section, we derive the electron-acoustic phonon coupling Hamiltonian from the deformation potential theory.
In principle, very similar expressions as the ones obtained for the optical phonons could be used.
In practice, those expressions do not clearly show how the e-ph coupling for acoustic phonons vanishes in the $q\to 0$ limit.
Consequently, we follow another approach, discussed in Ref.~[\onlinecite{mahan_book},\onlinecite{abrikosov_book}] to derive the leading-order electron-phonon coupling.

Since the acoustic phonon frequency vanishes at $q\to 0$, Ref.~[\onlinecite{abrikosov_book}] suggests to use the following operator instead of the operator $Q_\lambda(q)$~\cite{abrikosov_book}
\begin{align}
\tilde{Q}_\lambda(\mathbf{q}) =  \omega_\lambda(\mathbf{q})Q_\lambda(\mathbf{q}) = \sqrt{\frac{\hbar\omega_{\lambda}(\mathbf{q})}{2M }} \left(b_{\mathbf{q},\lambda}+b^\dagger_{-\mathbf{q},\lambda}\right).
\end{align}
In terms of $\tilde{Q}_\lambda(\mathbf{q})$,
Ref.~[\onlinecite{abrikosov_book}] further suggests that the coupling between long-wavelength acoustic phonons and electrons at valley is
\begin{align}
\label{eq:H_eph_abrikosov}
{\cal H}_{e-ph} = \sum_{\mathbf{k},\mathbf{q}}\sum_{\sigma\tau,\sigma'\tau'}\sum_{\lambda,\kappa} \tilde{g}_{\sigma\tau,\sigma'\tau'}^{\lambda,\kappa} \tilde{Q}_\lambda(\mathbf{q}) c_{\sigma\tau,\kappa}^\dagger(\mathbf{k}) c_{\sigma'\tau',\kappa}(\mathbf{k}-\mathbf{q}),
\end{align}
where $\tilde{g}_{\sigma\tau,\sigma'\tau'}^{\lambda,\kappa}$ are all independent of the phonon momentum $\mathbf{q}$, to leading order.
Now we transform Eq.~\eqref{eq:H_eph_abrikosov} back to the e-ph coupling in terms of $Q_\lambda(q)$
\begin{equation}
\mathcal{H}_{e-ph} = \sum_{{\bf k},{\bf q}}\sum_{\lambda,\kappa}\sum_{\sigma\tau,\sigma'\tau'} g^{\lambda,\kappa}_{\sigma\tau,\sigma'\tau'}(\mathbf{q}) Q_\lambda(\mathbf{q}) c_{\sigma\tau,\kappa}^\dagger ({\bf k}) c_{\sigma'\tau',\kappa}(\mathbf{k}-\mathbf{q})\ ,
\end{equation}
where Eq.~\eqref{eq:H_eph_abrikosov} suggests that the matrix $g^{\lambda,\kappa}$ must be of linear order of the $|\mathbf{q}|$.
Clearly, although the e-ph coupling matrix is $q-$independent for long-wavelength optical phonons, it is linear in ${\bf q}$ for acoustic phonons.
Nevertheless, Eq.~\eqref{eq:H_eph_abrikosov} does not take the direction of $\mathbf{q}$ into consideration, though the order of $\mathbf{q}$ is correct.
The direction of $\mathbf{q}$ does matter here, since the distortion effect breaks the four-fold rotational symmetry.
To include the direction of $\mathbf{q}$, we perform the symmetry analysis for $g^{\lambda,\kappa}$, while only keeping the linear order of $\mathbf{q}$.
After taking the combined symmetry $\mathcal{P}=C_{2x}\mathcal{T}$ and the Hermitian condition into account, all the matrix elements are purely imaginary.
Then, we arrive at
\begin{align}
\begin{split}
\mathbf{g}^{A_1,s} &= iq_x(g_0^a \tau_0 +  sg_2^a \tau_2 + g_3^a\tau_3) + iq_y(g_1^a\tau_1), \\
\mathbf{g}^{B_1,s} &= iq_y(g_0^a \tau_0 +  sg_2^a \tau_2 + g_3^a\tau_3) + iq_x(g_1^a\tau_1),
\end{split}
\end{align}
where $s$ is the valley index.
We emphasize that the pseudo-gauge field given by the acoustic phonons has the same form as the strain tensor, as reflected by linear $\bf q$ dependence of the e-ph couplings\cite{vozmediano2010}.

\section{Green's function for two coupled in-plane phonon modes}
\begin{figure}[!htbp]
	\centering
	\includegraphics[width=0.7\linewidth]{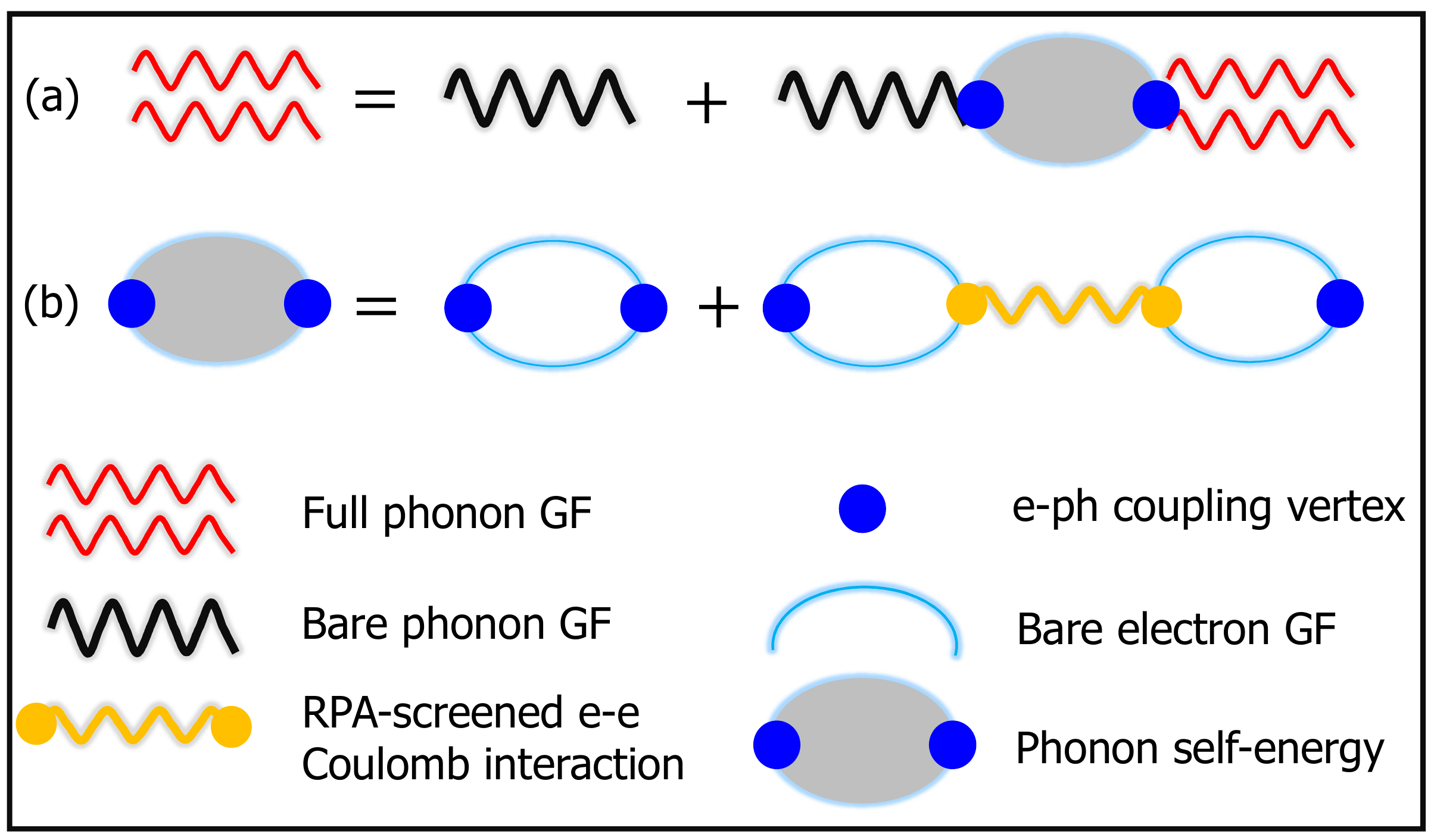}
	\caption{ (a) Dyson equation for the phonon Green's function (GF). (b) Equation for the phonon self-energy.
	}
	\label{sm_fig2}
\end{figure}

In this section, we discuss the phonon Green's function in the presence of electron-phonon (e-ph) and electron-electron (e-e) interactions.
The diagrammatic form of the Green's function is shown in Fig.~\ref{sm_fig2}, which follows Ref.~[\onlinecite{tse2008}] (we have reproduced the same results using a functional integral approach, in which electrons are integrated out with the aid of a Hubbard-Stratonovich transformation in order to obtain an effective action for phonons).
We will focus on the two optical in-plane phonons labeled by $\lambda=\{A_1,B_1\}$.
Then, the Matsubara Green's function for the phonons is defined as
\begin{align}\label{eq:D_def_gen}
D_{\lambda\lambda'}(\mathbf{q},\tau) = - \langle T_\tau A_{\mathbf{q},\lambda}(\tau) A_{-\mathbf{q},\lambda'}(0)\rangle,
\end{align}
where $A_{\mathbf{q},\lambda}=b_{\mathbf{q},\lambda}+b^\dagger_{-\mathbf{q},\lambda}$, $\tau$ is the imaginary time  and $T_\tau$ denotes the imaginary-time ordered product.
Note that the displacement operator is $Q_\lambda(\mathbf{q})= \sqrt{\frac{\hbar}{2M\omega_\lambda}} A_{\mathbf{q},\lambda}$.
In $(\mathbf{q},\omega)$ space,  we use
\begin{align}
D_{\lambda\lambda'}(\mathbf{q},iq_m) = \int_0^{\beta} e^{iq_m\tau} D_{\lambda\lambda'}(\mathbf{q},\tau) \, d\tau,
\end{align}
where $q_m=2\pi m/\beta$ with $\beta=1/k_BT$ the inverse of temperature and integer $m$.
Then, the bare (non-interacting) phonon Green's function is given by
\begin{align}
D_0(\mathbf{q},iq_m) = \text{Diag}\{D_{A_1A_1},D_{B_1B_1}\},
\end{align}
where $D_{\lambda\lambda}=2\omega_{\lambda}/((iq_m)^2-\omega_{\lambda}^2)$ and $\omega_{\lambda}$ is the bare frequency for the $\lambda$-phonon mode.
In the presence of electron-phonon interactions, the full Green's function is no longer diagonal in the $\{A_1,B_1\}$ basis.
In our calculation, we first ignore the $g_2$-term in the e-ph coupling Hamiltonian because it only renormalizes the Dirac electron's mass;  we discuss its effect to the off-diagonal phonon self-energy later (see Sec.~\ref{section-g2-self-energy}).
The full phonon Green's function is given by
\begin{align}\label{eq:D_D0_Sigma}
D(\mathbf{q},iq_m) = \left[ 1 - D_0(\mathbf{q},iq_m)  \Sigma(\mathbf{q},iq_m) \right]^{-1}D_0(\mathbf{q},iq_m),
\end{align}
where $\Sigma(\mathbf{q},iq_m)= \Sigma^p(\mathbf{q},iq_m) + \Sigma^e(\mathbf{q},iq_m)$
is the phonon self-energy due to bare e-ph interactions ($\Sigma^p$) and the screened e-e interaction ($\Sigma^e$).
The phonon self-energy can be represented as a $2\times 2$ matrix in  $\{A_1,B_1\}$ space. Its diagonal entries are
\begin{align}\label{sm-eq-diagonal-sf}
\begin{split}
\Sigma_{AA}^p &= \frac{\hbar}{2M\omega_A}(\Pi_{00}^{p} + \Pi_{33}^{p} + \Pi_{03}^{p} + \Pi_{30}^{p}) \text{ and } \Sigma_{BB}^p=\frac{\hbar}{2M\omega_B}\Pi_{11}^{p}, \\
\Sigma_{AA}^e &= \frac{\hbar}{2M\omega_A}(\Pi_{00}^{e} + \Pi_{33}^{e} + \Pi_{03}^{e} + \Pi_{30}^{e}) \text{ and } \Sigma_{BB}^e=\frac{\hbar}{2M\omega_B}\Pi_{11}^{e}.
\end{split}
\end{align}
Likewise, the off-diagonal self-energies are
\begin{align}\label{sm-eq-off-diagonal-sf}
\begin{split}
\Sigma_{AB}^p &=\frac{\hbar}{2M\sqrt{\omega_A\omega_B}}(\Pi_{01}^{p}+\Pi_{31}^{p}),\\
\Sigma_{AB}^e &=\frac{\hbar}{2M\sqrt{\omega_A\omega_B}}(\Pi_{01}^{e}+\Pi_{31}^{e}),
\end{split}
\end{align}
with $\Sigma_{BA} ({\bf q}, \omega) = \Sigma_{AB}(-{\bf q}, - \omega)$.
Again, we will take $\hbar=1$ convention in the following calculations for simplicity.
Here, the components $\Pi^{p}_{ij}({\bf q},iq_m)$ with $i,j=0,1,3$ are given by
\begin{align}\label{sm-eq-self-energy-Pi-ij}
\Pi_{ij}^{p}(\mathbf{q},i q_m)=\frac{g_i^o g_j^o}{\beta}\sum_{s,\mathbf{k},\omega_n} \text{tr}\lbrack \tau_i G_s(\mathbf{k},i\omega_n) \tau_j G_s(\mathbf{k}',i\omega_n+iq_m) \rbrack,
\end{align}
where $\mathbf{k}'=\mathbf{k}+\mathbf{q}$, $s$ labels the two valleys, $G_s(\mathbf{k},i\omega_n)=[-i\omega_n-\mu + h_s(\mathbf{k})]^{-1}$ is the bare Matsubara Green's function of Dirac electrons at valley $K_s$, $\omega_n=(2n+1)/\beta$ is a fermionic Matsubara frequency and $q_m=2m/\beta$ is a bosonic Matsubara frequency.
The retarded phonon self-energy $\Sigma({\bf q},\omega)$ is obtained by analytical continuation $i q_m \to \omega+ i 0^+$.

Similarly,
\begin{align}\label{sm-eq-self-energy-Pi-ij-e}
\Pi_{ij}^e(\mathbf{q},iq_m)=\frac{g_i^o g_j^o}{\beta^2 \mathcal{V}} \left( \sum_{s,\mathbf{k},\omega_n} \text{tr}\lbrack \tau_i G_s(\mathbf{k},i\omega_n) \tau_0 G_s(\mathbf{k}',i\omega_n+iq_m) \rbrack  V_q^{RPA}  \sum_{s,\mathbf{k},\omega_l'} \text{tr}\lbrack \tau_0 G_s(\mathbf{k},i \omega_l') \tau_j G_s(\mathbf{k}',i\omega_l'+i q_m) \rbrack \right),
\end{align}
where $V_q^{RPA}=V_q/(1-V_q\Pi_0^e)$  is the RPA-screened e-e interaction, with $V_q=e^2/(\epsilon_\infty q^2)$ the ``bare'' Coulomb potential ($\epsilon_\infty$ denotes the screening from high-energy electronic bands that are not included in Eq.~\eqref{sm-eq-h0-electron-2-by-2}).
$\mathcal{V}$ is the 3D volume.
Also, the electronic polarization function $\Pi_0^e$ is defined as
\begin{align}
\Pi^{e}_0(\mathbf{q},iq_m) = \int _0^{2\pi/a_z} \frac{dk_z}{2\pi} \left\{ \frac{1}{\beta \mathcal{A}} \sum_{s,\mathbf{k},\omega_n} \text{Tr}\lbrack \tau_0 G(\mathbf{k},\omega_n) \tau_0 G(\mathbf{k}+\mathbf{q},\omega_n+q_m) \rbrack \right\}  .
\end{align}
which will be calculated in the next subsection.
Here $\mathcal{A}$ is the area of the 2D Sb layer, the electric polarization is defined in 3D with a sum over $\mathbf{k}$ in 2D and one integral along $k_z$, and $a_z$ is the z-direction lattice constant.
Since the BaMnSb is just a layered material, and we assume the inter-layer coupling is zero.

The units for the quantities in the above expressions are summarized as $[Q]=[nm],[ c_\mathbf{k}]=[1], [V]=[nm^3], [a_z]=[nm], [\frac{\hbar}{2M\omega_A}] = [nm^2], [g]=[meV\cdot nm^{-1}], [G]=[meV^{-1}], [\beta]=[meV^{-1}],[ V_q]=[meV\cdot nm^{3}], [\sum_{k}]=[1], [\sum_{\omega}]=[1]$, then $[\Pi^{p}]=[\Pi^{e}]=[meV\cdot nm^{-2}]$, $[\Sigma]=[meV]$, and $[\Pi^{e}_0]=[meV^{-1}\cdot nm^{-3}]$.
Moreover, $[2M\omega_A]=[nm^{-2}]$ for $\hbar=e=1$ (this convention is used in the main text).
In the following sections, we take $[nm]=[meV^{-1}]$ to double-check the units for simplicity by taking $[v_0]=[1]$ ($v_0$ is Fermi velocity of Dirac fermions).

\subsection{The electronic polarization function for the insulating phase}
In this subsection, we calculate the contribution of low-energy electrons in BaMnSb$_2$ to the electronic polarization function $\Pi_0^e({\bf q}, i q_m)$.
We assume throughout that the chemical potential is inside the electron bandgap.
We calculate the contributions from the $K_+$ and $K_-$ valleys separately, and add them in the end.
The corresponding electronic Green's function at valley $\mathbf{K}_s$ (a $2\times 2$ matrix) is
\begin{align}
G_s(\mathbf{k},i\omega_n) = \left(-i\omega_n-\mu + h_s(\mathbf{k})\right)^{-1}= \frac{P_+(\mathbf{k})}{-i\omega_n -\mu+E_{\mathbf{k}}} +   \frac{P_-(\mathbf{k})}{-i\omega_n -\mu - E_{\mathbf{k}}},
\end{align}
where $P_{\pm}(\mathbf{k}) = \frac{1}{2}\lbrack  1 \pm h_{s}({\bf k}) /E_{\mathbf{k}} \rbrack$ are projector operators and $E_{\mathbf{k}}=\sqrt{v_0^2(k_x^2+k_y^2)+m^2}$.
We begin with contributions from electrons at $K_+$ valley to the electric polarization function,
\begin{align}
\frac{1}{\beta} \sum_{\mathbf{k}} \sum_{\omega_n} \text{Tr}\lbrack \tau_0 G(\mathbf{k},\omega_n) \tau_0 G(\mathbf{k}+\mathbf{q},\omega_n+q_m) \rbrack  .
\end{align}
First, we recognize that
\begin{align}
\tau_0 G(\mathbf{k},\omega_n) \tau_0 G(\mathbf{k}+\mathbf{q},\omega_n+q_m) = \sum_{\alpha\beta} f_{\alpha\beta}(\mathbf{k},\mathbf{q},\omega_n,q_m) P_{\alpha}(\mathbf{k}) P_{\beta}(\mathbf{k}+\mathbf{q}),
\end{align}
where $\alpha,\beta=\{+,-\}$ and $f_{\alpha\beta}(\mathbf{k},\mathbf{q},\omega_n,q_m)=\frac{1}{-i\omega_n -\mu + \alpha E_{\mathbf{k}}}\frac{1}{-i(\omega_n+q_m) -\mu + \beta E_{\mathbf{k}+\mathbf{q}}}$.
Then, the summation over Matsubara frequencies gives
\begin{align}
\begin{split}
k_BT \sum_{\omega_n} f_{\alpha\beta}(\mathbf{k},\mathbf{q},\omega_n,q_m)
&= \frac{n_F(\mu-\alpha E_{\mathbf{k}}) - n_F(\mu-\beta E_{\mathbf{k}+\mathbf{q}})}{-iq_m+\beta E_{\mathbf{k}+\mathbf{q}} - \alpha E_{\mathbf{k}}},
\end{split}
\end{align}
where $n_F(x) = 1/(1+e^{x/k_BT})$ is the Fermi distribution function so that $n_F(x>0)=0$ and $n_F(x<0)=1$ at zero temperature. For the trace part, we have
\begin{align}
\text{Tr}[P_\alpha(\mathbf{k}) P_\beta(\mathbf{k}+\mathbf{q})]
&=\frac{1}{2} + \frac{1}{2}\alpha\beta \left\lbrack \frac{E_{\mathbf{k}}}{E_{\mathbf{k}+\mathbf{q}}} + \frac{v_0^2(k_xq_x+k_yq_y)}{E_{\mathbf{k}}E_{\mathbf{k}+\mathbf{q}}} \right\rbrack.
\end{align}
The contribution from electrons at the $K_-$ valley can be computed in the same way, simply by replacing the projection $P_{\alpha}$ with $P_{-\alpha}$.
Therefore, summing over the contribution from both valleys, we obtain
\begin{align}
\Pi^{e}_0(\mathbf{q},iq_m) = \frac{1}{a_z\mathcal{A}}\sum_{\mathbf{k}}\sum_{\alpha\beta} \frac{n_F(\mu-\alpha E_{\mathbf{k}}) - n_F(\mu-\beta E_{\mathbf{k}+\mathbf{q}})}{-iq_m+\beta E_{\mathbf{k}+\mathbf{q}} - \alpha E_{\mathbf{k}}}  \left\{ 1 + \alpha\beta \left\lbrack \frac{E_{\mathbf{k}}}{E_{\mathbf{k}+\mathbf{q}}} + \frac{v_0^2(k_xq_x+k_yq_y)}{E_{\mathbf{k}}E_{\mathbf{k}+\mathbf{q}}} \right\rbrack\right\}.
\end{align}
In the $m=0$ limit, the term inside the curly brackets tends to $1+\alpha\beta\cos(\theta_{\mathbf{k},\mathbf{k}'})$,
where $\theta_{\mathbf{k},\mathbf{k}'}$ is the angle between $\mathbf{k}$ and $\mathbf{k}'$.
In that limit, our expression for $\Pi^{e}_0$ reproduces that of graphene \cite{Hwang2007}.

In the zero-temperature limit and when the chemical potential is inside the bandgap, only interband transitions $\alpha\neq \beta$ contribute to $\Pi^e_0$.
In the static limit, these contributions amount to a renormalization of $\epsilon_\infty$ in the screened Coulomb interaction.
More precisely, we obtain\cite{Pyatkovskiy2009}
\begin{align}
\Pi^{e}_0(\mathbf{q},\omega=0)\approx -q^2/(6a_z\pi\vert m_0\vert),
\end{align}
here $a_z$ is the z direction lattice constant.
which is valid when we take $v_0q\ll m_0$ limit.
Thus, the contribution from Dirac fermions to the static dielectric function is
\begin{align}\label{sm-eq-epsilon0-rpa}
\epsilon(\mathbf{q})=1 - V_q\Pi_0^e(\mathbf{q},\omega=0)\approx 1 +  e^2 / (6a_z\pi\epsilon_\infty \vert m_0\vert) \triangleq \epsilon_{RPA},
\end{align}
where $V_q= e^2/(\epsilon_\infty q^2)$ is the Coulomb potential.
We check the unit of $[V_q]=[meV^{-2}]$ and $[\Pi_0^e(\mathbf{q},iq_m)]=[meV^2]$, so that $\epsilon_{RPA}$ is a dimensionless parameter.
Thus, Eq.~\eqref{sm-eq-epsilon0-rpa} gives rise to the the RPA-screened Coulomb potential for the insulating electron system.

\subsection{The diagonal phonon self-energies $\Sigma_{AA}$ and $\Sigma_{BB}$}
In this subsection, we discuss the diagonal phonon self-energies, $\Sigma_{\lambda\lambda}$,  for $\lambda=\{A_1,B_1\}$. These quantities were defined in Eq.~\eqref{sm-eq-diagonal-sf}.
We first calculate all the necessary elements of $\Pi^p_{ij}$, with $i,j=0,1,3$, following Eq.~\eqref{sm-eq-self-energy-Pi-ij}.
The calculation is similar to that of the polarization function described above.
The explicit expressions are as follows:
\begin{align}
\Pi_{00}^{p}(\mathbf{q},iq_m) & = (g_0^o)^2\sum_{\mathbf{k}}\sum_{\alpha\beta} \frac{n_F(\mu-\alpha E_\mathbf{k}) - n_F(\mu-\beta E_{k+\mathbf{q}})}{-iq_m+\beta E_{\mathbf{k}+\mathbf{q}} - \alpha E_\mathbf{k}}  \left\lbrack 1 + \alpha\beta \left\lbrack \frac{E_\mathbf{k}}{E_{\mathbf{k}+\mathbf{q}}} + \frac{v_0^2(k_xq_x+k_yq_y)}{E_\mathbf{k}E_{\mathbf{k}+\mathbf{q}}} \right\rbrack \right\rbrack,\\
\Pi_{33}^{p}(\mathbf{q},iq_m) &= (g_3^o)^2\sum_{\mathbf{k}}\sum_{\alpha\beta} \frac{n_F(\mu-\alpha E_\mathbf{k}) - n_F(\mu-\beta E_{\mathbf{k}+\mathbf{q}})}{-iq_m+\beta E_{\mathbf{k}+\mathbf{q}} - \alpha E_\mathbf{k}}  \left\lbrack 1 + \alpha\beta
\frac{v_0^2(k_y^2-k_x^2)+v_0^2(k_yq_y-k_xq_x)-m_0^2}{E_\mathbf{k}E_{\mathbf{k}+\mathbf{q}}} \right\rbrack, \\
\Pi_{03}^{p}(\mathbf{q},iq_m) &= g_0^og_3^o \sum_{\mathbf{k}}\sum_{\alpha\beta} \frac{n_F(\mu-\alpha E_\mathbf{k}) - n_F(\mu-\beta E_{\mathbf{k}+\mathbf{q}})}{-iq_m+\beta E_{\mathbf{k}+\mathbf{q}} - \alpha E_\mathbf{k}} \left\lbrack i\alpha\beta\frac{v_0q_xm_0}{E_\mathbf{k}E_{\mathbf{k}+\mathbf{q}}} \right\rbrack = -\Pi_{30}^{p}(\mathbf{q},iq_m),  \label{sm-eq-sf-pi03} \\
\Pi_{11}^{p}(\mathbf{q},iq_m) &= (g_1^o)^2\sum_{\mathbf{k}}\sum_{\alpha\beta} \frac{n_F(\mu-\alpha E_\mathbf{k}) - n_F(\mu-\beta E_{\mathbf{k}+\mathbf{q}})}{-iq_m+\beta E_{\mathbf{k}+\mathbf{q}} - \alpha E_\mathbf{k}}  \left\lbrack 1 - \alpha\beta
\frac{v_0^2(k_y^2-k_x^2)+v_0^2(k_yq_y-k_xq_x)+m_0^2}{E_\mathbf{k}E_{\mathbf{k}+\mathbf{q}}} \right\rbrack, \\
\Pi_{01}^{p}(\mathbf{q},iq_m) &= g_0^og_1^o \sum_{\mathbf{k}}\sum_{\alpha\beta} \frac{n_F(\mu-\alpha E_\mathbf{k}) - n_F(\mu-\beta E_{\mathbf{k}+\mathbf{q}})}{-iq_m+\beta E_{\mathbf{k}+\mathbf{q}} - \alpha E_\mathbf{k}} \left\lbrack -i\alpha\beta\frac{v_0q_ym_0}{E_\mathbf{k}E_{\mathbf{k}+\mathbf{q}}} \right\rbrack = -\Pi_{10}^{p}(\mathbf{q},iq_m).
\label{sm-eq-sf-pi01}
\end{align}
In the insulating regime ($\mu$ inside the electron band gap), only interband transitions ($\alpha\neq\beta$) contribute at zero temperature.
For latter reference, when $|\omega|< 2|m_0|$ (where $\omega$ is a real frequency), the retarded response functions $\Pi_{ii}^p({\bf q},\omega)$ are purely real ($i=0,1,3$), while the retarded response functions $\Pi^p_{01}({\bf q},\omega)$ and $\Pi^p_{03}({\bf q},\omega)$ are purely imaginary.

\subsubsection{Diagonal self-energies in the absence of electron-electron interactions}
We first consider only the e-ph interaction.
Since $\Pi_{03}^p+\Pi_{30}^p=0$, the diagonal self-energies are given by $\Sigma_{AA}^p=\frac{1}{2M\omega_A}[\Pi_{00}^{p} + \Pi_{33}^{p}] $ and $\Sigma_{BB}^p=\frac{1}{2M\omega_B}\Pi_{11}^{p}$.
Before calculations, we double-check the unit of $[\Sigma_{AA,AB,BB}]=[meV]$, since $[M]=[meV]$, $[\omega_{A,B}]=[meV]$ and $[\Pi^p]=[meV^3]$.
In the insulating regime and if $|\omega|<2 |m_0|$, these self-energies are purely real at zero temperature.

\subsubsection{Contribution of electron-electron interactions to the diagonal self-energies}
Now, we incorporate the e-e interactions, which contribute to the diagonal self-energies as follows:
\begin{align}
\Sigma_{AA}^e(\mathbf{q},iq_m) &= \frac{1}{2M\omega_A}\left\lbrack \frac{1}{(g_0^o)^2 \mathcal{V}} \Pi_{00}^{p}(\mathbf{q},iq_m) V_q^{RPA} \Pi_{00}^{p}(\mathbf{q},iq_m) +  \frac{1}{(g_0^o)^2 \mathcal{V}} \Pi_{30}^{p}(\mathbf{q},iq_m)V_q\Pi_{03}^{p}(\mathbf{q},iq_m) \right\rbrack , \\
\Sigma_{BB}^e(\mathbf{q},iq_m) &= \frac{1}{2M\omega_B}\left\lbrack \frac{1}{(g_0^o)^2\mathcal{V}} \Pi_{10}^{p}(\mathbf{q},iq_m)V_q^{RPA} \Pi_{01}^{p}(\mathbf{q},iq_m) \right\rbrack.
\end{align}
In the insulating regime and if $|\omega|<2 |m_0|$, these self-energies are purely real at zero temperature.

\subsection{The off-diagonal phonon self-energy $\Sigma_{AB}$}
Next, let us consider the off-diagonal self-energy defined in Eq.~\eqref{sm-eq-off-diagonal-sf}, which couples the two optical phonons at a finite phonon momentum and generates the phonon helicity.
To compute this self-energy, we need the following quantities:
\begin{align}
\Pi_{01}^{p}(\mathbf{q},iq_m) &=  g_0^og_1^o \sum_{\mathbf{k}}\sum_{\alpha\beta} \frac{n_F(\mu-\alpha E_\mathbf{k}) - n_F(\mu-\beta E_{\mathbf{k}+\mathbf{q}})}{-iq_m+\beta E_{\mathbf{k}+\mathbf{q}} - \alpha E_\mathbf{k}} \left\lbrack -i\alpha\beta\frac{v_0q_ym_0}{E_\mathbf{k}E_{\mathbf{k}+\mathbf{q}}} \right\rbrack = -\Pi_{10}^{p}(\mathbf{q},iq_m), \\
\Pi_{31}^{p}(\mathbf{q},iq_m) &= g_3^og_1^o \sum_{\mathbf{k}} \frac{n_F(\mu-\alpha E_\mathbf{k}) - n_F(\mu-\beta E_{\mathbf{k}+\mathbf{q}})}{-iq_m+\beta E_{\mathbf{k}+\mathbf{q}} - \alpha E_\mathbf{k}}\times\left\lbrack \alpha\beta\frac{v_0^2(k_x(k_y+q_y)+k_y(k_x+q_x)}{E_\mathbf{k}E_{\mathbf{k}+\mathbf{q}}} \right\rbrack.
\end{align}

\subsubsection{Off-diagonal self-energy in the absence of electron-electron interactions}
Hereafter, we focus on $\Sigma_{AB}^p$ computed for ${\bf q}=(q_x,0)$ or ${\bf q}=(0,q_y)$.
These special cases will allow us to gain some analytical understanding on the phonon helicity.
The trace part for $\Pi_{31}^{p}(q,iq_m)$ is an odd function of $k_x$ or $k_y$, and thus
\begin{align}
\Pi_{31}^{p}(q_x,q_y=0,iq_m) &= g_3^og_1^o \sum_{\mathbf{k},\omega_n}
\frac{n_F(\mu-\alpha E_\mathbf{k}) - n_F(\mu-\beta E_{\mathbf{k}+\mathbf{q}})}{-iq_m+\beta E_{\mathbf{k}+\mathbf{q}} - \alpha E_\mathbf{k}}
\times\left\lbrack \alpha\beta\frac{v_0^2(2k_x+q_x)}{E_kE_{k+q}} \right\rbrack \times k_y =0,\\
\Pi_{31}^{p}(q_x=0,q_y,iq_m) &= g_3^og_1^o \sum_{\mathbf{k},\omega_n}
\frac{n_F(\mu-\alpha E_\mathbf{k}) - n_F(\mu-\beta E_{\mathbf{k}+\mathbf{q}})}{-iq_m+\beta E_{\mathbf{k}+\mathbf{q}} - \alpha E_\mathbf{k}}\times\left\lbrack \alpha\beta\frac{v_0^2(2k_y+q_y)}{E_kE_{k+q}} \right\rbrack \times k_x =0.
\end{align}
It follows that
\begin{align}\label{sm-eq-sf-piab-1}
\begin{split}
\Sigma_{AB}^p(q_x,q_y=0,iq_m) &= \frac{1}{2M\sqrt{\omega_A\omega_B}} \left[ \Pi_{31}^{p}(q_x,q_y=0,iq_m) + \Pi_{01}^{p}(q_x,q_y=0,iq_m) \right] \\
&=  \frac{1}{2M\sqrt{\omega_A\omega_B}}  \Pi_{01}^{p}(q_x,q_y=0,iq_m), \\
\Sigma_{AB}^p(q_x=0,q_y,iq_m) &= \frac{1}{2M\sqrt{\omega_A\omega_B}} \left[ \Pi_{31}^{p}(q_x=0,q_y,iq_m) + \Pi_{01}^{p}(q_x=0,q_y,iq_m) \right] \\
&=  \frac{1}{2M\sqrt{\omega_A\omega_B}}  \Pi_{01}^{p}(q_x=0,q_y,iq_m),
\end{split}
\end{align}

\subsubsection{Contribution of electron-electron interactions to the off-diagonal self-energy}
Next, we consider the additional correction to the off-diagonal phonon coming from e-e interactions, which is given by
\begin{align}
\Sigma_{AB}^e(\mathbf{q},iq_m) = \frac{1}{2M\sqrt{\omega_A\omega_B}} \left\lbrack \frac{1}{(g_0^o)^2\mathcal{V}} \Pi_{00}^{p}(\mathbf{q},iq_m) V_q^{RPA} \Pi_{01}^{p}(\mathbf{q},iq_m)  +  \frac{1}{(g_0^o)^2\mathcal{V}} \Pi_{30}^{p}(\mathbf{q},iq_m)V_q^{RPA} \Pi_{01}^{p}(\mathbf{q},iq_m) \right\rbrack.
\end{align}
Let us consider these two terms one-by-one.
From Eqs.~\eqref{sm-eq-sf-pi03} and \eqref{sm-eq-sf-pi01}, we obtain
\begin{align}
\Pi_{30}^{p}(\mathbf{q},iq_m)\sim q_x \text{ and } \Pi_{01}^{p}(\mathbf{q},iq_m)\sim q_y,
\end{align}
to linear order in ${\bf q}$. As a result, one of these two terms must vanish when ${\bf q}$ is along the $x$ or $y$ axis.
Then, it simplifies the off-diagonal self energy as
\begin{align}\label{sm-eq-sf-piab-2}
\Sigma_{AB}^e(\mathbf{q},iq_m) = \frac{1}{2M\sqrt{\omega_A\omega_B}} \left\lbrack \frac{1}{(g_0^o)^2\mathcal{V}} \Pi_{00}^{p}(\mathbf{q},iq_m) V_q^{RPA} \Pi_{01}^{p}(\mathbf{q},iq_m) \right\rbrack,
\end{align}
with ${\bf q}=(q_x,0)$ or ${\bf q}=(0,q_y)$.

Adding the results from Eqs.~\eqref{sm-eq-sf-piab-1} and \eqref{sm-eq-sf-piab-2}, the total off-diagonal self-energy becomes
\begin{align}
\begin{split}
\Sigma_{AB}(\mathbf{q},iq_m) &
= \left[ 1+\frac{1}{(g_0^o)^2\mathcal{V}} \Pi_{00}^{p}(\mathbf{q},iq_m) V_q^{RPA}  \right] \frac{1}{2M\sqrt{\omega_A\omega_B}} \Pi_{01}^{p}(\mathbf{q},iq_m), \\
&=\left[ 1+\Pi_{0}^{e}(\mathbf{q},iq_m) V_q^{RPA}  \right] \frac{1}{2M\sqrt{\omega_A\omega_B}} \Pi_{01}^{p}(\mathbf{q},iq_m) \\
&= \frac{1}{\epsilon(\mathbf{q})} \frac{1}{2M\sqrt{\omega_A\omega_B}}\Pi_{01}^{p}(\mathbf{q},iq_m) ,
\end{split}
\end{align}
where the RPA dielectric function $\epsilon(\mathbf{q})$ is given by Eq.~\eqref{sm-eq-epsilon0-rpa}.
In other words, the influence of e-e interactions in the off-diagonal phonon energy amounts to renormalizing the e-ph coupling parameters as
\begin{align}
g_0^o g_1^o \to \frac{1}{\epsilon(\mathbf{q})} g_0^o g_1^o \approx \frac{ g_0^o g_1^o}{\epsilon_{RPA}}.
\end{align}
In sum, after renormalizing the e-ph couplings, we have
\begin{align}
\Sigma_{AB}(\mathbf{q},iq_m) = \frac{1}{2M\sqrt{\omega_A\omega_B}}[\Pi_{01}^{p}(\mathbf{q},iq_m)],
\end{align}
with ${\bf q}=(q_x,0)$ or ${\bf q}=(0,q_y)$.
This is the result used in the main text to determine the phonon helicity.

\subsubsection{Symmetry constraints on the off-diagonal self-energy}
Now, let us focus on the symmetry constraints on the off-diagonal self-energy. The symmetry operators include the time-reversal symmetry and the two-fold rotation symmetry.
In this subsection, we work with real frequencies.
\begin{itemize}
	\item[(1.)] Time-reversal symmetry $\mathcal{T}$ imposes the following constraints on the electron Green's function:
	\begin{align}
	\mathcal{T} G_{\pm}(k_x,k_y,\omega) \mathcal{T}^{-1} = G_{\mp}^\ast(-k_x,-k_y,\omega).
	\end{align}
	It follows that
	\begin{align}
	\begin{split}
	\Pi_{i1}^p(\mathbf{q},\omega) &= \sum_{s}x_i(s) g_i^o g_1^o \frac{1}{\beta}\sum_{\mathbf{k},\omega'} \text{tr}\lbrack \tau_i G_s(\mathbf{k},\omega') \tau_1 G_s(\mathbf{k}',\omega'') \rbrack, \\
	&= \left(\sum_{s}x_i(s) g_i^o g_1^o \frac{1}{\beta}\sum_{\mathbf{k},\omega'} \text{tr}\lbrack \tau_i^\ast G_s^\ast(\mathbf{k},\omega') \tau_1^\ast G_s^\ast(\mathbf{k}',\omega'') \rbrack \right)^\ast, \\
	&= \left(\sum_{s} x_i(s) g_i^o g_1^o \frac{1}{\beta}\sum_{-\mathbf{k},\omega'} \text{tr}\lbrack \tau_i^\ast G_{-s}(-\mathbf{k},\omega') \tau_1 G_{-s}(-\mathbf{k}',\omega'') \rbrack \right)^\ast = \left(\Pi_{i1}^p(-\mathbf{q},\omega)\right)^\ast,
	\end{split}
	\end{align}
	where $x(s)=s\delta_{i,2}+\delta_{i,0}+\delta_{i,3}$ with $i=0,2,3$, we replace $s\to -s$ in the last line, and we use
	$x(-s)\tau_i^\ast = x(s)\tau_i$.
	Therefore, we have
	\begin{align}
	\Sigma_{AB}(\mathbf{q},\omega) = \Sigma_{AB}^\ast(-\mathbf{q},\omega).
	\end{align}
	\item[(2.)] The two-fold rotation symmetry $C_{2x}$ imposes the following constraint on the Green's function:
	\begin{align}
	C_{2x} G_{\pm}(k_x,k_y,\omega) C_{2x}^{-1} = G_{\mp}(k_x,-k_y,\omega).
	\end{align}
	Along the lines of the analysis above,
	\begin{align}
	\begin{split}
	\Pi_{i1}^p(q_x,q_y,\omega) &= \sum_{s}x_i(s) g_i^o g_1^o \frac{1}{\beta}\sum_{\mathbf{k},\omega'} \text{tr}\lbrack \tau_i G_s(\mathbf{k},\omega') \tau_1 G_s(\mathbf{k}',\omega'') \rbrack, \\
	&= \sum_{s}x_i(s) g_i^o g_1^o \frac{1}{\beta}\sum_{\mathbf{k},\omega'} \text{tr}\lbrack C_{2x} \tau_i C_{2x}^{-1} C_{2x} G_s(\mathbf{k},\omega') C_{2x}^{-1} C_{2x} \tau_1 C_{2x}^{-1} C_{2x} G_s(\mathbf{k}',\omega'') C_{2x}^{-1} \rbrack, \\
	&= -\sum_{s}x_i(s) g_i^o g_1^o \frac{1}{\beta}\sum_{\mathbf{k},\omega'} \text{tr}\lbrack C_{2x} \tau_i C_{2x}^{-1} G_{-s}(k_x,-k_y,\omega') \tau_1  G_{-s}(k_x',-k_y',\omega'') \rbrack ,\\
	&= -\Pi_{i1}^p(q_x,-q_y,\omega),
	\end{split}
	\end{align}
	where we used $x_i(s) C_{2x} \tau_i C_{2x}^{-1} = x_i(-s)\tau_i$ with $i=0,2,3$.
	Then we arrive at
	\begin{align}
	\Sigma_{AB}(q_x,q_y,\omega) = -\Sigma_{AB}(q_x,-q_y,\omega).
	\end{align}
\end{itemize}
Based on the above two symmetries, we find $\Sigma_{AB}({\bf q},\omega=0) \approx iq_y  (\partial_{q_y}\text{Im}[\Sigma_{AB}])_{{\bf q}=0} $ as shown in the main text.

\subsection{Berry curvature contribution to the off-diagonal self-energy in the \(\omega = 0\) limit}
In this section, we prove that the off-diagonal self-energy $\Sigma_{AB}(q_y,\omega)$ is proportional to TKNN invariant in the $q_y\to0$ and $\omega=0$ limit.
As discussed in the last subsection, $\Sigma_{AB}(q_y,\omega=0)$ is linearly proportional to $q_y$ due to the TR symmetry and $C_{2x}$ rotational symmetry.
Next, we demonstrate that the coefficient of the linear term is given by
\begin{align}\label{sm-equ-self-energy-N3}
a_0\frac{\partial}{\partial_{q_y}} \text{Im}[\Sigma_{AB}(q_y,\omega=0)] \vert_{q_y=0} = N_3,
\end{align}
where $a_0=\frac{2Mv_0\sqrt{\omega_A\omega_B}}{g_0g_1\mathcal{A}}$, and the right hand side is given by	
\begin{align}
N_3 =
\frac{\epsilon^{\alpha\beta\gamma}}{6} \text{tr}\int_{-\infty}^{\infty} d\omega' \int \frac{d^2 \mathbf{k}}{(2\pi)^2} G^{-1} \partial_{k_\alpha}G G^{-1} \partial_{k_\beta}G G^{-1} \partial_{k_\gamma}G
\end{align}
is the TKNN invariant and $\alpha,\beta,\gamma\in\{i\omega', k_x,k_y \}$.
Without loss of generality, we set $\hbar=1$.
To prove the above equation, let us begin with $a_0\Sigma_{AB}(q_y)$ with a symmetric form
\begin{align}
\frac{1}{\beta \mathcal{A}} \sum_{\mathbf{k}} \sum_{\omega_n} \text{Tr}\lbrack \tau_0 G(\mathbf{k}-\frac{\bf q}{2},\omega_n) \tau_x G(\mathbf{k}+\frac{\bf q}{2},\omega_n+q_m) \rbrack.
\end{align}
After taking the partial differential with the $q_y\to0$ limit, we find that
\begin{align}
\partial_{q_y}G(\mathbf{k}\pm\mathbf{\frac{q}{2}},\omega_n)\vert_{q_y=0} = \pm\frac{1}{2}\partial_{k_y}G(\mathbf{k},\omega_n).
\end{align}
Therefore, the left hand side of Eq.~\eqref{sm-equ-self-energy-N3} becomes
\begin{align}\label{sm-equ-left-hand-1}
\frac{1}{4}\text{tr}\int_{-\infty}^{\infty} d\omega' \int \frac{d^2 k}{(2\pi)^2} \left( -\partial_{k_y}G \tau_xG + G\tau_x\partial_{k_y}G \right).
\end{align}
For a single Dirac Hamiltonian of the type $h({\bf k}) = v_0(k_x\tau_x + k_y\tau_y) + m\tau_z$, with velocity $v_0 $, we have
\begin{align}\label{sm-equ-left-hand-2}
\frac{1}{v_0}\partial_{k_x} G^{-1} = -\tau_x, \; \partial_{i\omega} G^{-1} = \tau_0.
\end{align}
Substituting Eq.~\eqref{sm-equ-left-hand-2} into Eq.~\eqref{sm-equ-left-hand-1}, we find
\begin{align}\label{sm-equ-left-hand-3}
a_0\frac{\partial}{\partial_{q_y}} \text{Im}[\Sigma_{AB}(q_y,\omega=0)] \vert_{q_y=0} = \frac{1}{2}\text{tr}\int_{-\infty}^{\infty} d\omega' \int \frac{d^2 k}{(2\pi)^2} \Big{(} -(\partial_{k_y} G) (\partial_{k_x} G^{-1}) G (\partial_{i\omega'} G^{-1})
+ G(\partial_{k_x} G^{-1}) (\partial_{k_y} G) (\partial_{i\omega'} G^{-1}) \Big{)}.
\end{align}
By using the identity
\begin{align}
G\partial_{\alpha} G^{-1} = -(\partial_{\alpha} G) G^{-1}
\end{align}
with $\alpha=\{i\omega',k_x,k_y\}$,
we can simplify the Eq.~\eqref{sm-equ-left-hand-3} as
\begin{align}\label{sm-equ-left-hand-4}
\frac{1}{2}\text{tr}\int_{-\infty}^{\infty} d\omega' \int \frac{d^2 k}{(2\pi)^2} \Big{(} (G^{-1} \partial_{k_x} G) (G^{-1}\partial_{k_y} G) (G^{-1}\partial_{i\omega'} G)
-(G^{-1} \partial_{k_y} G) (G^{-1}\partial_{k_x} G) (G^{-1}\partial_{i\omega'} G) \Big{)} = N_3.
\end{align}
where the cyclic permutation for the trace is used.
This concludes the proof of Eq.~(\ref{sm-equ-self-energy-N3}).

\section{The additional self-energy correction due to the $g_2^o$ e-ph coupling vertex} \label{sm-section-g2-self-energy}
Up until now, we have studied the contribution to the phonon self-energy originating from $g_0^o$, $g_1^o$ and $g_3^o$; the corresponding electron-phonon interactions behave as pseudo-gauge fields for electrons.
In doing so, we have ignored the effect of the $g_2^o$ term, which renormalizes the Dirac mass of the electrons. One may ask how this term will influence our main results concerning the electronic Berry curvature contribution to the phonon helicity. To address this question, we write the correction from the $g_2^o$ e-ph coupling to the off-diagonal phonon self-energy as
\begin{align}
\Delta\Sigma_{AB} = \Delta\Sigma^p_{AB} + \Delta\Sigma^e_{AB},
\end{align}
where $\Delta\Sigma^p_{AB} = \frac{1}{2M\sqrt{\omega_A\omega_B}} \Pi_{21}^p$, $\Delta\Sigma^e_{AB}=\frac{1}{2M\sqrt{\omega_A\omega_B}}\frac{1}{(g_0^o)^2\mathcal{V}}\Pi_{20}^p\times V_q^{RPA} \times \Pi_{01}^p $ and
\begin{align}
\Pi_{01}^{p}(\mathbf{q},iq_m) &=  g_0^og_1^o \sum_{\mathbf{k}}\sum_{\alpha\beta} \frac{n_F(\mu-\alpha E_\mathbf{k}) - n_F(\mu-\beta E_{\mathbf{k}+\mathbf{q}})}{-iq_m+\beta E_{\mathbf{k}+\mathbf{q}} - \alpha E_\mathbf{k}} \left\lbrack -i\alpha\beta\frac{v_0q_ym_0}{E_\mathbf{k}E_{\mathbf{k}+\mathbf{q}}} \right\rbrack , \\
\Pi_{21}^p(\mathbf{q},iq_m) &= g_2^og_1^o \sum_{\mathbf{k}}\sum_{\alpha\beta} \frac{n_F(\mu-\alpha E_\mathbf{k}) - n_F(\mu-\beta E_{\mathbf{k}+\mathbf{q}})}{-iq_m+\beta E_{\mathbf{k}+\mathbf{q}} - \alpha E_\mathbf{k}} \left[ i\alpha\frac{v_0k_y}{E_{\mathbf{k}}} -i\beta\frac{v_0(k_y+q_y)}{E_{\mathbf{k}+\mathbf{q} }} \right],\\
\Pi_{20}^p(\mathbf{q},iq_m) &= g_2^og_0^o \sum_{\mathbf{k}}\sum_{\alpha\beta} \frac{n_F(\mu-\alpha E_\mathbf{k}) - n_F(\mu-\beta E_{\mathbf{k}+\mathbf{q}})}{-iq_m+\beta E_{\mathbf{k}+\mathbf{q}} - \alpha E_\mathbf{k}} \left[ \alpha \frac{m}{E_{\mathbf{k} }}  +\beta \frac{m}{E_{ \mathbf{k} + \mathbf{q}  } } \right].
\end{align}

\subsection{The self-energy correction $\Delta\Sigma^p_{AB}$ in the absence of electron-electron interactions}
In this subsection, let us first study the self-energy correction $\Delta\Sigma^p_{AB} = \frac{1}{2M	\sqrt{\omega_A\omega_B}} \Pi_{21}^p$. Because we focus on the insulating phase at zero temperature,
we can set $\mu=0$. Accordingly,
\begin{align}
n_F(-E_{\mathbf{k}}) = n_F(-E_{\mathbf{k}+\mathbf{q}})=1, \text{ and } n_F(E_{\mathbf{k}}) = n_F(E_{\mathbf{k}+\mathbf{q}})= 0,
\end{align}
which leads to
\begin{align}
\Pi_{21}^p(\mathbf{q},\omega) \approx g_2^og_1^o \times  i\omega \sum_{\mathbf{k}} \frac{2}{(E_{\mathbf{k}} + E_{\mathbf{k}+\mathbf{q}})^2}
\left[ \frac{v_0k_y}{E_{\mathbf{k}}} + \frac{v_0(k_y+q_y)}{E_{\mathbf{k}+\mathbf{q}}} \right].
\end{align}
Here, we have used $iq_m\to \omega + i0^+$ and we have kept only the leading order term in $\omega$.
We check that $\Pi_{21}^p$ satisfies the symmetry requirements with respect to time-reversal and the two-fold rotation $C_{2x}$,
\begin{align}
\Pi_{21}^p(q_x,q_y=0,\omega) =0, \quad \Pi_{21}^p(q_x=0,q_y,\omega) = -\Pi_{21}^p(q_x=0,-q_y,\omega), \quad \Pi_{21}^p(q_x,q_y,\omega) = \left( \Pi_{21}^p(-q_x,-q_y,\omega) \right)^\ast.
\end{align}
To compare with the $\Pi_{01}^p \propto iq_y$ term,
we evaluate $\Pi_{21}^p$ to leading order in $q_y$ and we focus on the results along the $q_y$-axis, i.e., $q_x=0$.
Performing a Taylor expansion on $q_y$ and keeping terms up to linear order, we arrive at
\begin{align}
\begin{split}
\Pi_{21}^p(q_x=0,q_y,\omega) &\approx g_2^og_1^o \times  i\omega\sum_{\mathbf{k}} \frac{1}{2E_{\mathbf{k}}^2}\left( 1-\frac{v_0^2k_yq_y}{E_{\mathbf{k}}^2} \right)  \left[ \frac{v_0k_y}{E_{\mathbf{k}}} + \frac{v_0q_y}{E_{\mathbf{k}}} + \frac{v_0k_y}{E_{\mathbf{k}}}\left( 1-\frac{v_0^2k_yq_y}{E_{\mathbf{k}}^2} \right) \right] \\
&\approx g_2^og_1^o \times  i\omega\sum_{\mathbf{k}} \left[ \frac{v_0q_y}{2E_{\mathbf{k}}^3}  +  \frac{v_0k_y}{2E_{\mathbf{k}}^3}\times\left( 2-3\frac{v_0^2k_yq_y}{E_{\mathbf{k}}^2}  \right)  \right]
= g_2^og_1^o \times  i\omega\sum_{\mathbf{k}} \left[ \frac{v_0q_y}{2E_{\mathbf{k}}^3}  - \frac{v_0k_y}{2E_{\mathbf{k}}^3}\frac{3v_0^2k_yq_y}{E_{\mathbf{k}}^2}  \right] \\
&= g_2^og_1^o \times  i\omega v_0q_y \sum_{\mathbf{k}} \left[ \frac{1}{2E_{\mathbf{k}}^3}  - \frac{3v_0^2k_y^2}{2E_{\mathbf{k}}^5}  \right]  = g_2^og_1^o \times  i\omega v_0q_y  \left( \frac{1}{2\vert m\vert v_0^2} - \frac{1}{2} \frac{3v_0^2}{2} \frac{2}{3\vert m\vert v_0^4} \right) = 0.
\end{split}
\end{align}
This indicates that the leading order term must be $q_y^3$, namely,
\begin{align}
\Pi_{21}^p(q_x=0,q_y,\omega) \propto g_2^og_1^o \times (i\omega q_y^3).
\end{align}
In sum, in the long-wavelength regime, the contribution of the $g_2$-term to the off-diagonal phonon self-energy is small compared to that from $\Pi_{01}^p$.

\subsection{Contribution of electron-electron interactions to the self-energy correction $\Delta\Sigma^e_{AB}$}
In this subsection, we study the self-energy correction $\Delta\Sigma^e_{AB}=\frac{1}{2M\sqrt{\omega_A\omega_B}}\frac{1}{(g_0^o)^2\mathcal{V}}\Pi_{20}^p\times V_q^{RPA} \times \Pi_{01}^p $, where $\Pi_{01}^p\sim iq_y$.
To do so, we only need to evaluate $\Pi_{20}^p$.
Setting $q_x=0$, a straightforward calculation gives
\begin{align}
\Pi_{20}^p (q_x=0,q_y,\omega) &\approx g_2^og_0^o \times m\omega  \sum_{\mathbf{k}}  \frac{2}{(E_{\mathbf{k}} + E_{\mathbf{k}+\mathbf{q}})^2}   \left[ \frac{1}{E_{\mathbf{k}}} - \frac{1}{E_{\mathbf{k}+\mathbf{q}}} \right],
\end{align}
By inspection,
we notice that $\Pi_{20}^p (q_x=0,q_y,\omega)$ is at least of order $q_y^2$.
Performing a Taylor expansion to order $q_y^2$, we arrive at
\begin{align}
\begin{split}
\Pi_{20}^p (q_x=0,q_y,\omega) &\approx g_2^og_0^o \times m\omega  \sum_{\mathbf{k}}  \frac{1}{2E_{\mathbf{k}}^3} \left[1 - \frac{v_0^2k_yq_y}{E_\mathbf{k}^2}  \right]  \left[   \frac{v_0^2k_yq_y}{E_\mathbf{k}^2}  + \frac{v_0^2q_y^2}{2{E_\mathbf{k}^2}} \left( 1-\frac{3v_0^2k_y^2}{E_\mathbf{k}^2}  \right)   \right], \\
&\approx g_2^og_0^o \times m\omega  \sum_{\mathbf{k}}   \frac{1}{2E_{\mathbf{k}}^3} \left[  \frac{v_0^2k_yq_y}{E_\mathbf{k}^2} + \frac{v_0^2q_y^2}{2E_\mathbf{k}^2}  \left( 1 - \frac{5v_0^2k_y^2}{E_\mathbf{k}^2}   \right) \right], \\
&\approx 	g_2^og_0^o \times m\omega \left(\frac{1}{4}v_0^2q_y^2\right) \times \left( 0 + \frac{1}{3\vert m\vert^3 v_0^2} - 5v_0^2\frac{1}{2}\frac{2}{15\vert m\vert^3 v_0^4} \right) =0.
\end{split}
\end{align}
This indicates that the leading order must be $q_y^4$, namely,
\begin{align}
\Pi_{20}^p (q_x=0,q_y,\omega) \sim g_2^og_0^o \times m\omega q_y^4.
\end{align}
Consequently,
\begin{align}
\Delta\Sigma^e_{AB}=\left(\frac{1}{g_0^o}\right)^2\Pi_{20}^p\times V_q^{RPA} \times \Pi_{01}^p \sim  g_2^og_1^o \times (i\omega q_y^3).
\end{align}
In sum, the full contribution of the $g_2^o$ e-ph coupling to the off-diagonal phonon self-energy scales as
\begin{align}
\Delta\Sigma_{AB} = \Delta\Sigma^p_{AB} + \Delta\Sigma^e_{AB} \sim \Delta\Sigma^p_{AB} \sim  g_2^og_1^o \times (i\omega q_y^3).
\end{align}
For long-wavelength phonons, this correction can be neglected compared to the $\Sigma_{AB} \propto iq_y$ contribution obtained from $g_0^o$ and $g_1^o$.

	\section{Field-theoretic derivation for the off-diagonal coupling between phonons}
	In this section, we show an alternative derivation for phonon hybridization from the field theory.
	Here we use the real time, which is related to the imaginary-time formalism in other sections by a Wick rotation.
	
	We begin by integrating out the 2D Dirac fermions~\cite{comtet2000aspects} in Eq.~\eqref{eq:pseudo-gauge}.
	This gives the following Chern-Simons term in the effective action:
	\begin{align}
	\mathcal{S}_{eff} = \sum_{s}\frac{\sigma_s}{2} \int dt \, d^2\mathbf{r} \, \epsilon^{\mu\nu\rho}  \widetilde{A}_{s,\mu} \partial_\nu \widetilde{A}_{s,\rho},
	\end{align}
	where the Einstein summation convention is adopted for the Greek letters ($\mu,\nu,\rho$), $s$ is the valley index,
	\eq{
		\sigma_s=\frac{s}{4\pi}  \sgn{m_0}
	} is the Hall conductance, and we have assumed zero temperature with the chemical potential inside the mass gap ($\hbar=e=1$ units are used).
	So we have $[\mathcal{S}_{eff}] =[1]$ due to $[t]=[meV^{-1}], [r]=[meV^{-1}], [\widetilde{A}_{s,\mu}]=[meV]$.
	The effective action can be split into three parts
	\eq{
		S_{eff}= S_{eff}^A+S_{eff}^{A-pse}+S_{eff}^{pse}\ ,
	}
	where $S_{eff}^A$ is the part solely for the $U(1)$ gauge field $A$, $S_{eff}^{A-pse}$ labels the interaction between the $U(1)$ gauge field and the pseudo-gauge field, and $S_{eff}^{pse}$ is solely for the pseudo-gauge field.
	
	Next, we apply the above equation to BaMnSb$_2$. For $S_{eff}^{pse}$, we can use the form of the pseudo-gauge field \eqnref{eq:A_pse} to derive the coupling between $A_1$ and $B_1$ optical phonons, resulting in
	\eq{
		S_{eff}^{pse}= \mathcal{A}T \sum_q \frac{\sgn{m_0}}{4\pi}\frac{g_0 g_1}{ v_0 } q_y\bsl{Q}^T_q \mat{ 0 & \ii \\ -\ii &0 } \bsl{Q}_{-q} + \text{diagonal part}\ ,
	}
	where $\bsl{Q}_q= (Q_{A_1,q},Q_{B_1,q})^T$ with $q=(\ii \omega,\bsl{q})$, and $T$ is redefined as the time interval in this section.
	Check the unit: $[\mathcal{A}]=[meV^{-2}], [T]=[meV^{-1}], [g_{0,1}]=[meV^2], [v_0]=[1], [q_y]=[meV], [\bsl{Q}_q]= [meV^{-1}]$, so that $[S_{eff}^{pse}]=[1]$.
	The diagonal parts will renormalize the phonon frequency as discussed in Sec.~(IV.B).
	Following the standard techniques in Ref.~[\onlinecite{Srednicki2007QFT}], we take functional derivations of the partition function to calculate the total phonon Green's function. 
	Here we are using the real-time Green's function, which is done by converting \eqnref{eq:D_def_gen} and \eqnref{eq:D_D0_Sigma} to the real time.
	Then, the off-diagonal term plays the role of self-energy corrections, which is given by
	\eq{\label{eq:Sigma_CS}
		\Sigma_{AB}= \mathcal{A} \frac{1}{2\pi}\frac{g_0g_1}{v_0}\frac{m_0}{\vert m_0\vert} q_y\frac{1}{2M\sqrt{\omega_A\omega_B}}\ ,
	}
	which coincides with the results in Sec.~(IV.D) and Eq.~(6) in the main text, as $N_3^+-N_3^-= \frac{1}{2\pi} m_0/|m_0|$.
	Nevertheless, the expression in Eq.~(\ref{eq:Sigma_CS}) does not include the screening of electron-phonon couplings due to electron-electron interactions; our objective here is to provide a simple and intuitive reason for why the electronic Chern number alters phonon properties and leads to phonon helicity. 
	Screening effects of Sec.\,IV can be recovered in the present formalism by keeping higher order terms in electromagnetic gauge fields in the perturbative expansion of the phonon effective action.
	In the Sec.\,V above, we have shown the effect of electron-phonon interaction by analyzing all the self-energy corrections due to e-e and e-ph interactions.
	
	On the other hand, $S_{eff}^{A-pse}$ reads
	\eq{
		S_{eff}^{A-pse}=\frac{\sgn{m_0}}{2\pi}\int d^3 x (\frac{g_1}{v_0}Q_{B_1} E_y-\frac{g_3}{v_0}Q_{A_1} E_x)= Q\sum_q \bsl{E}_q^T  \bsl{Q}_{-q} + ...\ ,
	}
	where $Q= \mathcal{V} T\frac{\sgn{m_0}(g_1-g_3)}{4\pi v_0}$, and ``..." corresponds to the anisotropic part.
	In this work, we neglect the anisotropic part, and then the full effective action for optical phonons, including $\mathcal{S}_{eff}^{A-pse}$ and $\mathcal{S}_{eff}^{pse}$, reads
	\begin{align}\label{eq-chern-simon-action}
	\mathcal{S}_{ph} &=\frac{1}{2}\sum_{\omega,\mathbf{q}}  \mathbf{Q}_q^{T}  \mathcal{M}(\mathbf{q},\omega) \mathbf{Q}_{-q} + \sum_{\omega,\mathbf{q}} Q\mathbf{E}_q^T \bsl{Q}_{-q},   \\
	\mathcal{M}(\mathbf{q},\omega) &=\left[\begin{array}{cc}
	\omega^2-\omega_A^2  &  \sqrt{\omega_A\omega_B}\Sigma_{AB} \\
	-  \sqrt{\omega_A\omega_B}\Sigma_{AB}  & \omega^2-\omega_B^2
	\end{array}\right],
	\end{align}
	where $\omega_A^2$ and $\omega_B^2$ include the diagonal phonon self-energies' corrections, and $\Sigma_{BA}({\bf q}, \omega) = \Sigma_{AB}(-{\bf q}, -\omega)$ is used.

\section{Phonon helicity for acoustic phonons}
The general electron-acoustic phonon coupling vertex is given by
\begin{align}
\begin{split}
\mathbf{g}^{a,A_1,s} &= iq_x(g_0^a \tau_0 +  sg_2^a \tau_2 + g_3^a\tau_3) + iq_y(g_1^a\tau_1), \\
\mathbf{g}^{a,B_1,s} &= iq_y(g_0^a \tau_0 +  sg_2^a \tau_2 + g_3^a\tau_3) + iq_x(g_1^a\tau_1).
\end{split}
\end{align}
The off-diagonal self-energy is similar to that discussed for optical phonons and is given by
\begin{align}
\Sigma_{AB}(\mathbf{q},iq_m) = \frac{1}{2M\sqrt{\omega_A^a\omega_B^a}}[\Pi_{01}^{p}(\mathbf{q},iq_m)],
\end{align}
since we only focus on the $q_y$ and $q_x$ directions in the discussion of phonon dynamics.
Different from the optical phonon with a constant frequency $\omega_A$ and $\omega_B$ for the limit ${\bf q}\rightarrow 0$, the acoustic phonons possess a linear dispersion, given by $\omega_\lambda^a=v_\lambda\vert q\vert$ for the $\lambda$-mode in the limit ${\bf q}\rightarrow 0$.

\begin{figure}[!htbp]
	\centering
	\includegraphics[width=0.7\linewidth]{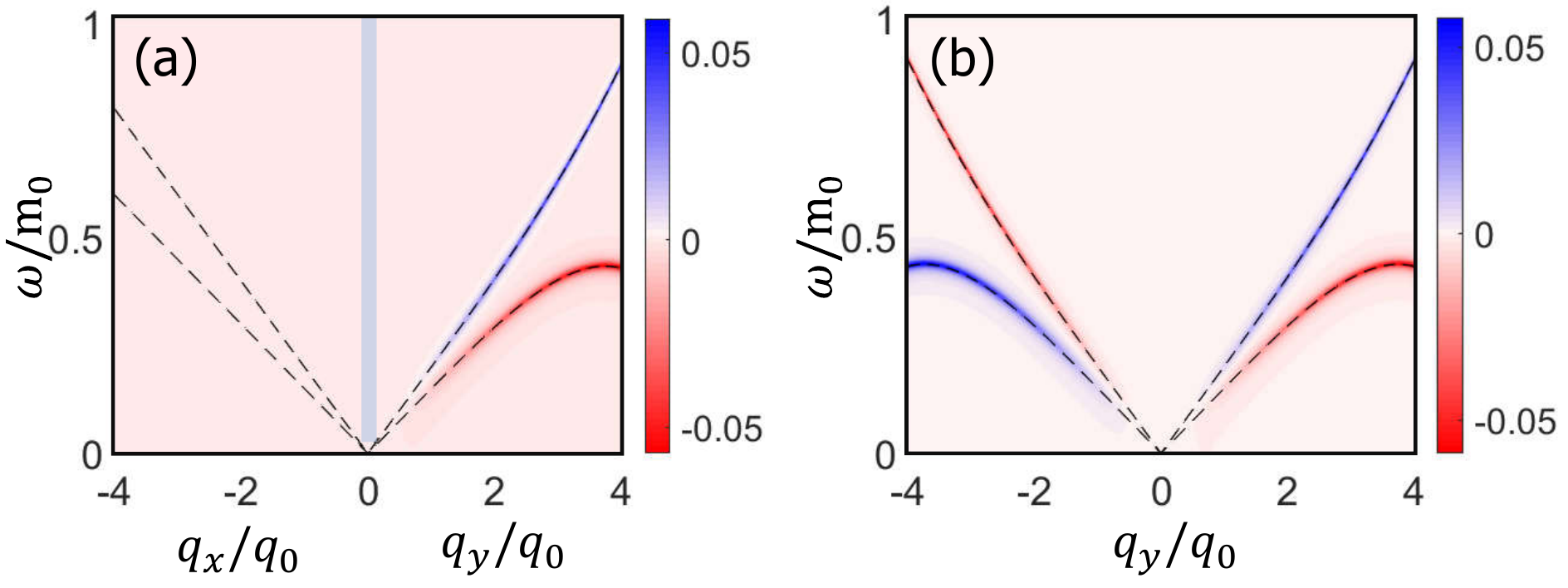}
	\caption{ (a) and (b) show the spectrum and circular polarization of optical phonons with $m_0=25$ meV.
		The phonon dispersions are labeled by black dashed lines, while the color represents the polarization.
		Parameters: $v_0=100$ meV$\cdot$nm, $\mu=0$, $v_{A_1}=15$ meV$\cdot$nm, $v_{B_1}=20$ meV$\cdot$nm, $g_0^a=80\sqrt{15}$ meV, $g_1^a=80\sqrt{20}$ meV. Here $q_0=m_0/v_0$ and $\Sigma_0= g_0g_1\mathcal{A}N_3^+ /(v_0M\sqrt{\omega_A\omega_B})$.
	}
	\label{sm_fig3}
\end{figure}

All the off diagonal self-energy results obtained from the electron-optical phonons coupling can be directly applied to the electron-acoustic phonon coupling.
The only difference is that $\Sigma_{AB}$ has $q_y^3$ dependence for acoustic phonons, instead of a linear $q_y$ dependence for optical phonons.
The numerical results are shown in Fig.~\ref{sm_fig3}, where the black lines depict the phonon dispersion including the correction from the off-diagonal self-energy. The colors in Fig.~\ref{sm_fig3} show non-vanishing circular phonon polarization or phonon angular momentum at a finite $q_y$. Similar to the optical phonons, we find helical structure of the phonon angular momentum as a function of the momentum ${\bf q}$ for acoustic phonons.

\section{The dynamical Kohn anomaly}
In this section, we analytically analyze the singularity and kinks for the off-diagonal phonon self-energy $\Sigma_{AB}(\mathbf{q},\omega) \sim \Pi_{01}^p(\mathbf{q},\omega)$ at finite $\omega$ and ${\bf q}$. Using the identity
\begin{align}
\frac{1}{x\pm i\eta} = \mathcal{P}\frac{1}{x} \mp i\pi \delta(x),
\end{align}
where $\mathcal{P}$ means the principal value, we obtain
\begin{align}
\frac{1}{-\omega-i\eta +\beta E_{\mathbf{k}+\mathbf{q}} - \alpha E_\mathbf{k}}
= \mathcal{P}\frac{1}{-\omega+\beta E_{\mathbf{k}+\mathbf{q}} - \alpha E_\mathbf{k}} + i\pi\delta(-\omega+\beta E_{\mathbf{k}+\mathbf{q}} - \alpha E_\mathbf{k}).
\end{align}
Thus, for the off-diagonal self-energy $\Sigma_{AB}(\mathbf{q},\omega) = \frac{1}{2M\sqrt{\omega_A\omega_B}}\Pi_{01}^{p}(\mathbf{q},\omega) $,
and the real component of $\Pi_{01}^{p}(\mathbf{q},\omega)$ is given by
\begin{align}
\text{Re}[\Pi_{01}^{p}(\mathbf{q},\omega)] = g_0 g_1\sum_{k}\sum_{\alpha\beta} \left( n_F(\mu-\alpha E_\mathbf{k}) - n_F(\mu-\beta E_{\mathbf{k}+\mathbf{q}})\right)
\left\lbrack \alpha\beta\frac{\pi v_0q_ym_0}{E_\mathbf{k}E_{\mathbf{k}+\mathbf{q}}} \right\rbrack \delta(-\omega+\beta E_{\mathbf{k}+\mathbf{q}} - \alpha E_\mathbf{k}).
\end{align}
For simplicity, we concentrate on the insulating regime ($0<=\mu<\vert m_0\vert <\omega/2$) at zero temperature.
With $n_F(\mu+E_\mathbf{k})=n_F(\mu+E_{\mathbf{k}'})=0$ and $n_F(\mu-E_\mathbf{k})=n_F(\mu-E_{\mathbf{k}'})=1$ ($\mathbf{k}'=\mathbf{k}+\mathbf{q}$), we obtain
\begin{align}
\frac{\text{Re}[\Pi_{01}^{p}(\mathbf{q},\omega)]}{g_0g_1\pi v_0m_0q_y} &= \sum_{\mathbf{k}}\frac{1}{E_\mathbf{k}E_{\mathbf{k}'}}\left( \delta(-\omega- E_{\mathbf{k}'} - E_\mathbf{k}) - \delta(-\omega + E_{\mathbf{k}'} + E_\mathbf{k})  \right) = -\sum_{\mathbf{k}}\frac{\delta(-\omega + E_{\mathbf{k}'} + E_\mathbf{k})}{E_\mathbf{k}E_{\mathbf{k}'}}.
\end{align}
In polar coordinates, we have
\begin{align}
\sum_{k} = \mathcal{A}\int \frac{d^2k}{(2\pi)^2} = \mathcal{A} \int_{0}^{\infty} kdk\int_{0}^{2\pi}d\theta,
\end{align}
where $\mathcal{A}$ is the area of the crystal in the $xy$ plane.
Next, we choose ${\bf q}=(0,q)$ with $q\ge0$, so that $\theta$-dependence only appears for the terms involving $k' = \sqrt{k^2+q^2+2kq\cos(\theta-\pi/2)}=\sqrt{k^2+q^2+2kq\sin	\theta}$.
Then, making a change of variables, we have
\begin{align}
\begin{split}
&\int_0^{2\pi}d\theta \to 2\int_{-\pi/2}^{\pi/2}d\theta
\to \int_{\vert k-q\vert}^{k+q} dk'\frac{4k'}{\sqrt{(2kq)^2-((k')^2-k^2-q^2)^2}}.
\end{split}
\end{align}
After this replacement, we obtain
\begin{align}
\frac{\text{Re}[\Pi_{01}^{p}(\mathbf{q},\omega)]}{g_0g_1\pi v_0m_0q_y} = -\mathcal{A} \int_0^\infty dk \int_{\vert k-q\vert}^{k+q} dk' g(k,k') \delta(-\omega + E_{\mathbf{k}'} + E_\mathbf{k}),
\end{align}
where $g(k,k')=\frac{4kk'}{\sqrt{(2kq)^2-((k')^2-k^2-q^2)^2}}$ is a continuous function and $E_k=\sqrt{k^2+m_0^2}$ with $v_0\equiv 1$.
With some algebra, this expression can be further simplified as
\begin{align}
\frac{\text{Re}[\Pi_{01}^{p}(\mathbf{q},\omega)]}{g_0g_1\pi v_0m_0q_y}
= -\mathcal{A} \int_0^{+\infty} dk \,  g(k,k_0) \theta(k_0-\vert k-q\vert)\theta(k+q-k_0),
\end{align}
where $\theta(...)$ is the step function, $k_0=\sqrt{(\omega-\sqrt{k^2+m_0^2})^2-m_0^2}$.
The integrand contains two step functions, which confine the range of the integration. Fix all these parameters here except $\omega$, we try to find if the range of integration changes by tunning $\omega$, which indicates a singularity/kink for this integration.
\begin{itemize}
	\item[(1.)] For the step function $\theta(k+q-k_0)$, the $k+q-k_0$ increase monotonically by $k$. And
	$\left( k+q-k_0\right)_{k=0} = q-k_0  $ can changes sign, once it happens, the range of the integration will change.
	\item[(2.)] For the step function $\theta(k_0-\vert k-q\vert)$,
	$k_0-\vert k-q\vert$ is a monotonically increasing function if $k<q$ and a monotonically decreasing function for $k>q$. Also, $\left( k_0-\vert k-q\vert\right)_{k=0}=k_0 - q$ could change sign.
\end{itemize}
Therefore, we find that $(k+q-k_0)_{k=0}=0$ gives rise to the singularity/kink, therefore, the kink locates at $q_0= \sqrt{(\omega-m_0)^2-m_0^2}$, shown in the main text.

\section{Spatial dispersion of the dielectric function}
In this section, we study the influence of the phonon helicity of optical phonons on the dielectric response of the material.
From the action for purely phonon modes (see Sec.~(VI)), 
the equation of motion for the phonon dynamics can be obtained my minimizing the effective action~\eqref{eq-chern-simon-action},
\begin{align}
\mathbf{Q}_\mathbf{q}  \mathcal{M}(\mathbf{q},\omega) + Q\mathbf{E}(\mathbf{q},\omega)=0.
\end{align}
The optical phonon can produce a macroscopic polarization $\mathbf{P}$, given by
$\mathbf{P}(\omega,q_y)=Q \mathbf{\Psi}_{q_y}(\omega) = \chi \mathbf{E}$ with the electric susceptibility $\chi$ to be a two-by-two matrix.
The complex dielectric function matrix is then given by
\begin{align}
\epsilon = \epsilon_0 I_{2\times2} + \chi  = \epsilon_0 I_{2\times2}
+ \frac{Q^2}{N(q_y,\omega)} \left[\begin{array}{cc}
\omega^2-\omega_B^2  &   C\\
-C  & \omega^2-\omega_A^2
\end{array}\right],
\end{align}
where $N(q_y,\omega)= (\omega^2-\omega_A^2)(\omega^2-\omega_B^2)-\omega_A\omega_B\Sigma_{AB}(q_y,\omega)$ and $C=\sqrt{\omega_A\omega_B}\Sigma_{AB}$.
Note that $C$ is purely imaginary.
Since the off-diagonal term is an odd function in $q_y$, one can check the dielectric function matrix satisfies the Onsager relation due to the time-reversal symmetry.
The off-diagonal component of the dielectric constant is of particular interest and is given by
$\epsilon_{xy}=i\sqrt{\omega_A\omega_B} \Sigma_{AB}(q_y,\omega)/N(q_y,\omega) \sim i\sqrt{\omega_A\omega_B}g_0g_1m_0q_y/(2\pi v_0\vert m_0\vert(\omega^2-\omega_A^2)(\omega^2-\omega_B^2))$ up to the lowest order in ${\bf q}$.
Along $q_x$ axis, the off-diagonal self-energy vanishes, indicating that the off-diagonal term of the dielectric function also becomes zero.
Therefore, one can see that the Berry curvature contribution to the optical phonon self-energy will also enter into the spatial dispersion (${\bf q}$ dependence) of the off-diagonal dielectric function through the optical phonon-light coupling. Since this term will have a resonance at the optical phonon frequency, it can be easily distinguished from other contribution through tuning the light frequency.
However, it may not be easy to measure this effect in optical absorption/reflectivity, because it vanishes when $q\to 0$ and the photon wave vectors corresponding to optical phonon frequencies are small.

\end{document}